\documentclass[authoryear]{elsarticle}
\usepackage{xcolor}

\begin{document}
\begin{frontmatter}

\title{Morphogenesis by coupled regulatory networks: 
Reliable control of positional information and 
proportion regulation}
\author[Epi,MPI]{Thimo Rohlf},
\author[Bremen]{Stefan Bornholdt}
\address[Epi]{Epigenomics Project, Genopole, Tour Evry 2, 523 Terrasses de l'Agora, F-91034 Evry cedex, France}
\address[MPI]{Max-Planck-Institute for Mathematics in 
the Sciences, Inselstrasse 22, D-04103 Leipzig, Germany}
\address[Bremen]{Institute for Theoretical Physics, 
University of Bremen, Otto-Hahn-Allee, D-28334 Bremen, 
Germany}
\date{\today}
\begin{abstract}
Based on a non-equilibrium mechanism for spatial 
pattern formation we study how position information 
can be controlled by locally coupled discrete dynamical 
networks, similar to gene regulation networks of cells 
in a developing multicellular organism. As an example 
we study the developmental problems of domain formation 
and proportion regulation in the presence of noise, as 
well as in the presence of cell flow. We find that 
networks that solve this task exhibit a hierarchical 
structure of information processing and are of similar 
complexity as developmental circuits of living cells. 
Proportion regulation is scalable with system size 
and leads to sharp, precisely localized boundaries 
of gene expression domains, even for large numbers 
of cells. A detailed analysis of noise-induced dynamics, 
using a mean-field approximation, shows that noise in 
gene expression states stabilizes (rather than disrupts) 
the spatial pattern in the presence of cell movements, 
both for stationary as well as growing systems.
Finally, we discuss how this mechanism could be realized 
in the highly dynamic environment of growing tissues 
in multi-cellular organisms.
\end{abstract}


\end{frontmatter}

\section{\label{intro}Introduction}
Understanding the molecular machinery that regulates development of multicellular
organisms is among the most fascinating problems of modern science. Today, a growing 
experimental record about the regulatory mechanisms involved in development is 
accumulating, in particular in well-studied model-organisms as, e.g., \emph{Drosophila} 
or \emph{Hydra} \citep{Technau2000,Bosch2003}. Still, the genomic details known today are not 
sufficient to derive dynamical models of developmental gene regulation processes in full detail.
Phenomenological models of developmental processes, on the other hand, are well established 
today. Pioneering work in this field was done by Turing, who in his seminal paper in 1952  
\citep{Turing1952} considered a purely physico-chemical origin of biological pattern formation. 
His theory is based on an instability in a system of coupled reaction-diffusion equations. 
In this type of model, for certain parameter choices, stochastic fluctuations in the initial 
conditions can lead to self-organization and maintenance of spatial patterns, e.g.\ 
concentration gradients or periodic patterns. This principle has been successfully 
applied to biological morphogenesis in numerous applications 
\citep{Gierer1972,Meinhardt2000}. However, as experiments make 
us wonder about the astonishingly high complexity of single regulating genes in 
development \citep{Bosch2002}, they also seem to suggest that diffusion models 
will not be able to capture all details of developmental regulation, and point at a complex 
network of regulating interactions instead. 

In theoretical work on pattern formation, 
both the crucial role of local, induction-like
phenomena in development \citep{Slack1993} and limitations of diffusion based mechanisms
in cellular environments \citep{Reilly1996} have lead to consideration of models that rely on local signal transfer
via membrane-bound receptors. A well-studied model in this context is {\em juxtacrine signaling} \citep{Wearing2000,Owen2000}.
Positive feed-back, combined with juxtacrine signaling, can lead to generation of spatial patterns
with wavelengths that extend over many cell lengths \citep{Owen2000}.
Further, it has been shown that a relay mechanism based on juxtacrine signaling can also
lead to travelling wave fronts, and hence provide an alternative mechanism
of long-range pattern regulation \citep{Monk1998}.
The emergence of {\em sharp spatial expression boundaries} of many genes
in development, besides other genes that exhibit more graded profiles, is notoriously hard
to explain in reaction-diffusion-based models. Recently it was shown in models of homeoprotein
intercellular transfer \citep{Kasatkin2007,Holcman2007} that restricted local diffusion of a morphogen regulating
its own expression can generate a morphogenetic gradient. When two of these gradients meet,
for certain parameter values a sharp boundary is created \citep{Kasatkin2007}.

The role of information processing in gene regulatory networks during development has 
entered the focus of theoretical research only recently. One pioneering study was published 
by \cite{Jackson1986}, who investigated the dynamics of spatial pattern
formation in a system of locally coupled, identical dynamical networks. In this model, gene 
regulatory dynamics is approximated by Boolean networks with a subset of nodes 
communicating not only with nodes in the (intracellular) network, but also with some 
nodes in the neighboring cells. Boolean networks are minimal models of information 
processing in network structures and have been discussed as models of gene regulation 
since the end of the 1960s \citep{Kauffman1969,Kauffman1993,Glass1973}. The model of Jackson et al.\ 
demonstrated the enormous pattern forming potential of local information processing.
More elaborate models that include a Boolean network description of cell-internal
gene regulatory networks, local inductive inter-cellular signals and a
discrete model of cell adhesion \citep{Hogeweg2000} point at a complex
interplay between regulatory dynamics, cell differentiation and morphogenesis.

Along similar lines, Salazar-Ciudad et al.\ introduced a gene network model based 
on continuous dynamics \citep{Salazar-Ciudad2000,Sole2002} and coupling their 
networks by direct contact induction. Interestingly, they observe a larger variety of 
spatial patterns than Turing-type models with diffusive morphogens, and find that 
patterns are less sensitive to initial conditions, with more time-independent (stationary) 
patterns. This matches well the intuition that networks of regulators have the potential for 
more general dynamical mechanisms than diffusion driven models. 
Furthermore, dynamical models of regulatory networks that control basic stages
in development, e.g. the segment polarity network in {\em Drosophila} embryos \citep{Dassow2000},
have shown that developmental modules are extremely robust against large parameter variations;
\cite{Albert2003} even showed that the topology of regulatory interactions alone in a Boolean
network model is sufficient to correctly predict both wild type and mutant patterns generated
by  the segment polarity network. Considering the temporal succession of
regulatory dynamics rather than spatial patterns,
similar results were obtained for other gene regulatory networks important
for cell development, e.g. the cell cycle network of different yeast species \citep{Li2004,Davidich2007}.
The fact that, in many instances, simple discrete dynamical network models
are sufficient to capture essential properties of developmental
dynamics, suggests that information-transfer-based processes controlled
by the topology of regulatory interactions both within and between cells
are very important for the extreme robustness and reliability
observed in development despite considerable noise and
large rearrangements of cell ensembles due to cell proliferation and -movements.
In this paper, we follow this new paradigm of interacting networks in pattern 
formation and in particular consider information-transfer-based processes.

We start with a particular problem of position-dependent gene activation,
as motivated from similar observations in {\em Hydra}. This animal 
is one of the most basal metazoa and exhibits extremely
precise regulation of expression boundaries under continuous cell movements.
Also, it has remarkable properties to regenerate {\em de novo} after
dissociation of cells, and to regulate its body proportions during growth.
We introduce a novel, two-level theoretical approach to model 
pattern formation problems of this type: First, a coarse-grained description in a
deterministic cellular automata model is developed, which then is extended to
a detailed model based on locally coupled, discrete dynamical networks.
We show that this deterministic model explains both de-novo
pattern formation after randomization of the pattern, and proportion
regulation of gene activity domains. A threshold network
model is derived which yields an upper estimate of the 
complexity of the regulatory module needed to solve
the pattern formation task.
Next,
model dynamics is studied under noise and cell movements, and solved
analytically in a mean-field approximation. It is shown
that local, stochastic changes in gene expression states
do not disrupt the spatial pattern, but contribute to
its stabilization in the presence of cell flow
by production of traveling domain boundaries ("quasi-particles")
that coordinate global positional information,
bearing some similarity to traveling waves found
in models based on juxtacrine signaling \citep{Monk1998}.
This suggests
an interesting, new mechanism for reliable morphogenetic control,
that might well apply to different types of tissues with high
demands on regeneration. It is shown that the mechanism also
works in growing tissues, explains pattern restoration
after cutting the tissue in half, and is robust against
noise in the detection of the body axis direction.
Last, potential applications of the model are discussed.

\section{Motivation and model}

Let us first introduce and define the morphogenetic problem that we will use as a 
motivation for our novel pattern formation model. 

\subsection{Proportion regulation in Hydra}\label{propreg_sec}
\begin{figure}[htb]
\begin{center}
\resizebox{70mm}{!}{\includegraphics{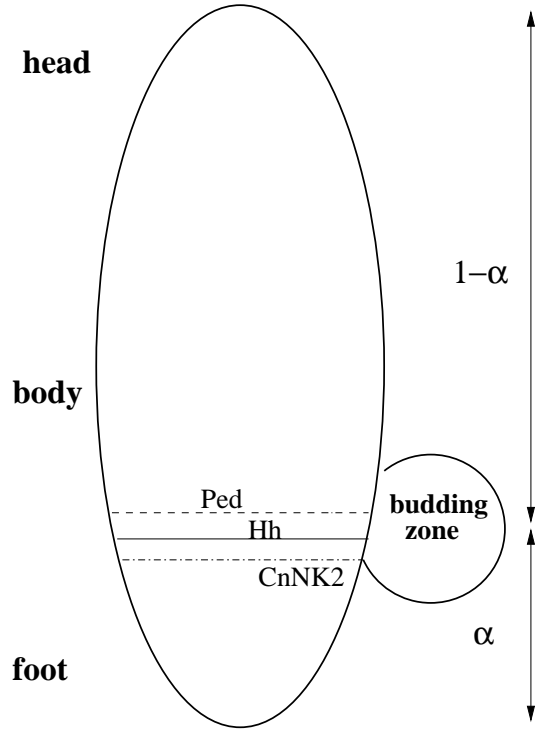}}
\end{center}
\caption{\small Gene expression domains in \emph{Hydra}, here for the example of the ``foot'' 
genes Pedibin (Ped) and Cn-NK2, and Hedgehog (Hh). Ped and Hh expression are bounded towards
the body region of the animal; while Ped exhibits a graded decay in the budding region,
Hh exhibits a sharp boundary. The relative position of the budding region and the associated expression boundaries, 
given by the ratio $\alpha/(1-\alpha)$, is independent of the absolute size of the animal (proportion regulation). Details
are explained in the text.}
\label{hydraped}
\end{figure}

A classical model organism for studies of position dependent gene activation is the fresh 
water polyp \emph{Hydra}, which has three distinct body regions - a head with mouth and 
tentacles, a body column and a foot region. The positions of these regions are accurately 
regulated along the body axis (Fig.\ \ref{hydraped}).

Hydra also has the capacity to reproduce asexually by exporting surplus cells into
{\em buds}; again, the position along the body axis where budding occurs is precisely regulated
at about two-thirds of the distance from the head to the foot \citep{Schiliro1999}. A number of genes
are involved in regulation of the foot region and the budding zone; for example,
both \emph{Pedibin} and \emph{CN-NK2} are only expressed in the foot region and turned
off approximately in the budding region \citep{Thomsen2004}. While the CN-NK2 expression
domain exhibits a rather graded decay in the budding region \citep{Grens1996}, it has been observed
that, for example, {\em Hedgehog (Hh)} is turned off {\em precisely} just below the budding region
with a sharp boundary \citep{Kaloulis2000}. The relative 
position of the budding region and of the
gene expression domains below it is almost independent of the animal's size, i.e.\ the ratio 
$\alpha/(1-\alpha)$ (as denoted in Fig.\ \ref{hydraped}) is almost invariant under changes 
of body size. As the Pedibin/CN-NK2 system presumably plays an important role 
in determining the foot region, this invariance appears to be an essential prerequisite for 
maintaining the correct body proportions (proportion regulation) and to establish the 
head-foot polarity. Very precise regulation
over a 10-fold size range has also been reported for the head-body proportion in Hydra,
with a value close to 1/3 \citep{Bode1980}.
Such precise regulation of position information and body proportions is a quite general problem 
in biological development \citep{Wolpert1969}.   

An interesting problem is how the specific properties of this regulation 
can be achieved by a small network of regulatory genes and if so, whether local communication 
between the cells (networks) is sufficient. This basic question is the central motivation for the 
present study. In particular, we consider the simplified problem of regulating one domain 
as, for example, the foot region versus the rest of the body. We consider this as a one 
dimensional problem as first approximation to the well-defined head-foot-axis in \emph{Hydra}.  
We should note, however, that we do not intend to model in detail the regulatory mechanisms
underlying \emph{Hydra} pattern formation, and rather take the observations made for this
basic metazoon as an inspiration to introduce and study a simple, generic model of pattern
formation.

\begin{figure}[htb]
\begin{center}
\resizebox{85mm}{!}{\includegraphics{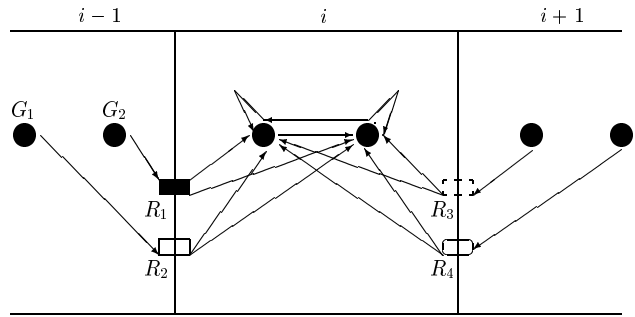}}
\end{center}
\caption{\small Diagram showing the interaction structure of the minimal network needed to solve 
the asymmetric expression task. For the sake of clarity, intracellular interactions between the two 
genes $G_1$ and $G_2$ are shown only for cell $i$, and likewise outgoing intercellular signals 
from the two genes two the neighbor cells $i-1$ and $i+1$ were left out. The transcription factors 
produced by gene $G_1$ and $G_2$ in cell $i-1$ couple to the receptor systems $R_1$ and $R_2$, 
respectively, whereas in cell $i+1$ the transcription factors produced by these genes couple to the 
receptor systems $R_3$ and $R_4$ (biased signaling). In cell $i$, the receptors release factors 
which regulate the activity of $G_1$ and $G_2$. }
\label{web1}
\end{figure}

Developmental processes exhibit an astonishing robustness. This often includes the ability of 
de novo pattern formation, e.g., to regenerate a
{\em Hydra} even after complete dissociation of the cell ensemble in a centrifuge \citep{Gierer1972b}. 
Further, they are robust in the face of a steady cell flux: \emph{Hydra} cells constantly move from 
the central body region along the body axis towards 
the top and bottom, where they differentiate into the respective cell types according to their position 
on the head-foot axis. The global pattern of gene activity is maintained in this dynamic environment.  
Let us take these observations as a starting point for a detailed study how the interplay of noise-induced 
regulatory dynamics and cell flow may stabilize a developmental system.

\subsection{One dimensional cellular automata: Definitions}
 
We here undertake a three-step approach to find a genetic network model that solves the pattern formation 
problem outlined above. In the first step, we summarize the properties of the cellular automata model introduced in 
\citep{Rohlf2005}. Cellular automata as dynamical systems discrete in time and state space 
are known to display a wide variety of complex patterns \citep{Wolfram1983,Wolfram1984,Wolfram1984a} and are 
capable of solving complex computational tasks, including universal computation.
We searched for solutions (i.e. rule tables which solve the problem) by aid of a genetic algorithm 
(for details, see Appendix). Candidate solutions have to fulfill four demands: Their update dynamics has 
to generate a spatial pattern which 1) obeys a predefined scaling ratio $\alpha/(1-\alpha)$, 2) is independent 
of the initial condition chosen at random, 3) is independent of the system size (i.e.\ the number of cells 
$N_C$) and 4) is stationary (a fixed point). In the second step, this cellular automata rule table is 
translated into (spatially coupled) Boolean networks, using binary coding of the cellular automata states. 
The logical structure of the obtained network is reduced to a minimal form, 
and then, in step three, translated into a threshold network. 

To define a model system that performs the pattern formation task of domain self-organization 
\citep{Rohlf2005}, consider a one-dimensional cellular automaton with parallel update
\citep{Wolfram1984}. $N_C$ cells are arranged on a one-dimensional lattice, and each cell is labeled 
uniquely with an index $i \in \{0,1,...,N_C-1\}$.
Each cell can take $n$ possible states $\sigma_{i} \in \{0,1,..,n\}$. The state $\sigma_i(t)$ of cell $i$
is a function of its own state $\sigma_i(t-1)$ and of its neighbor's states  $\sigma_{i-1}(t-1)$  and
 $\sigma_{i+1}(t-1)$ at time $t-1$, i.e. 
\begin{equation}\label{cadef}
\sigma_i(t) = f \left[ \sigma_{i-1}(t-1),  \sigma_i(t-1), \sigma_{i+1}(t-1) \right] 
\end{equation}
with $f: \{0,1,...,n\}^3  \mapsto \{0,1,...,n\}$ (a cellular automaton with {\em neighborhood 3}).
At the system boundaries, we set $\sigma_{-1} = \sigma_{N_C} = const. = 0$. Other choices,
e.g.\ asymmetric boundaries with cell update depending only on the inner neighbor cell, lead
to similar results. 
The state evolution of course strongly depends on the choice of $f$: for a three-state cellular automaton 
($n = 3$), there are $3^{27} \approx 7.626\cdot 10^{12}$ possible update rules, each of which
has a unique set of dynamical attractors. As we will show in the results section, $n=3$ is the minimal
number of states necessary to solve the pattern formation problem formulated above.

Now we can formulate the problem we intend to solve as follows: Find a set $\cal F$ of functions (update rules) 
which, given an initial vector $\vec{\sigma} = (\sigma_0,..., \sigma_{N_c-1})$ sampled randomly from
the set of all possible state vectors, within $T$ update steps 
evolves the system's dynamics to a fixed point attractor with the property:
\begin{equation}\label{probdef}
\vec{\sigma}^* :=  \left\{  \begin{array}{c} \sigma_i = 2   \quad if \quad i < [\alpha\cdot N_C] \\ 
                                                \sigma_i \ne 2 \quad if \quad i \ge [\alpha\cdot N_C] \end{array}\right. 
\end{equation}
where $[x]$ yields the largest integer value smaller or equal to the argument $x$ (this is needed since
the product $\alpha\cdot N_C$ may lead to non-integer values).
The scaling parameter $\alpha$ may take any value $0 < \alpha < 1$. For simplicity,
it is fixed here to $\alpha = 0.3$. Notice that $\alpha$ does not depend on $N_C$, i.e.\ we are looking for a set of solutions
where the ratio of the domain sizes $r:= \alpha / (1-\alpha)$ is {\em constant under changes of the system size},
as motivated in section \ref{propreg_sec} by similar observations of {\em proportion regulation} in developing 
multi-cellular organisms.
This clearly is a non-trivial task when only local information transfer is allowed. 
The ratio $r$ is a {\em global property} of the system, which has to emerge from purely local (next neighbor) interactions
between the cell's states.

\subsection{Pattern formation by locally coupled Boolean networks}

One can now take a step further towards biological systems, by transferring the 
dynamics we found for a cellular automata chain onto cells in a line that communicate 
with each other, similar to biological cells. Identifying different dynamical states with 
(differentiated) cell types \citep{Kauffman1969} and assuming 
that all model cells have an identical network of regulators inside, each of them capable to reproduce 
the rules of a cellular automaton by means of a dynamical coupling between 
subsets of regulators in direct neighbor cells,
we obtain a model mimicking basic properties of a 
biological genetic network in development. 

Cellular automata rule tables can easily be formalized as logical (Boolean) update tables, e.g., for $n=3$, 
two internal nodes 
with states $\sigma_1^i, \sigma_2^i \in \{0,1\}$
can be used for binary coding of the cell states (where $i$ labels the cell
position). One then
has
\begin{equation}\label{booldef}
(\sigma_{1,2}^i(t), \sigma_{2}^i(t))  = f_{1,2}\sigma_{1,2}^{i-1}(t-1), \sigma_{1,2}^i(t-1), \sigma_{1,2}^{i+1}(t-1))
\end{equation}
with $f_{1,2}: \{0,1\}^{6} \mapsto \{0,1\}^2$.
The so obtained rule tables are, by application of Boolean logic,
transformed into a minimized \emph{conjunctive normal form}, which only makes use of the the
three logical operators NOT, AND and OR with a minimal number of AND operations. This is a rather 
realistic approximation for gene regulatory networks, as the AND operation is more difficult to realize on the basis
of interactions between transcription factors. Other logical
functions as, e.g., XOR, are even harder to realize biochemically \citep{Davidson2001}. 
However, we should notice again that it is not our intention to develop a specific or detailed
biochemical model for the proposed pattern formation problem, but rather to find the simplest possible
network model that reproduces the basic phenomenology. Nonetheless, a model of this type
may well serve as a starting point for subsequent, more sophisticated models.
The basic structure of the constructed network and a possible biological interpretation is shown in Fig. \ref{web1}.
Concordant with the spirit we followed so far, we assume that symmetry of signal transmission is broken on a
\emph{local-cell scale}, without specifying this in detail; several self-organizing
mechanisms are conceivable in cellular systems, e.g.,
local chemical gradients \citep{Gurdon2001} or anisotropic distribution of receptors on cell membranes \citep{Galle2002}. 
In Fig.\ 2, for the sake of simplicity, the specific example of a biased distribution
of signal-transmitting receptors was chosen.

\begin{figure}[htb]
\begin{center}
\resizebox{85mm}{!}{\includegraphics{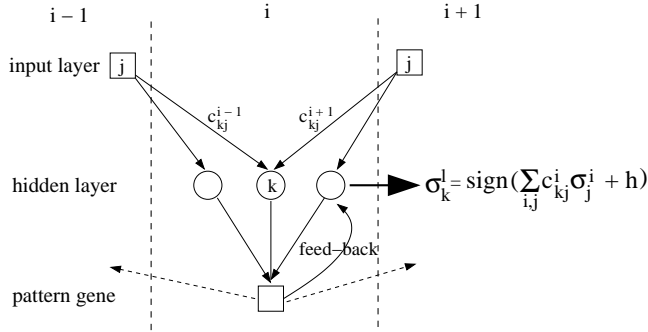}}
\end{center}
\caption{\small Schematic sketch of a threshold network controlling spatial gene activity
patterns. Signals from genes in direct neighbor cells constitute the \emph{input layer} of the
network. In the \emph{hidden layer}, this information is processed, with logical functions
implemented as weighted sums of the inputs. The output of the genes in this layer then controls the 
\emph{pattern genes}. Notice that there is feed-back to the hidden layer, as well as to
the neighbor cells (dashed arrows). }
\label{thresscheme}
\end{figure}

\subsection{The simplest dynamics: locally coupled threshold networks}\label{tn_method_subsec}

Perhaps the simplest dynamical model for transcriptional regulation networks are threshold networks, 
a subset of Boolean networks, where logical functions are modeled by weighted sums of the 
nodes' input states plus a threshold $h$ \citep{Kuerten1988a}. They
have proven to be valuable tools to address questions associated to the dynamics and 
evolution of gene regulatory networks 
\citep{Wagner1994,Bornholdt2000,Bornholdt2000a,Rohlf2002,Rohlf2004a}.

Any Boolean network can be coded as a dynamical threshold 
network with suitable thresholds assigned to each network node.
For the system of coupled networks discussed in this paper, 
this network contains minimally three hierarchies of information 
processing (``input layer'': signals from the genes in the
neighbor cells at time $t-1$, ``hidden layer'': 
logical processing of these signals, ``output layer'': 
states of the two ``pattern genes'' $\sigma_1^i$ and $\sigma_2^i$ 
in cell $i$ at time $t$ (see Fig. \ref{thresscheme} 
for a schematic sketch of the system structure)
\footnote{Notice that this structure is quite similar to a 
feed-forward neural network, however, in our system there 
is regulatory feed-back from the output-layer to the 'hidden' layer.}.
The genes' states now may take values $\sigma_i = \pm 1$, 
and likewise for the interaction weights one has $c_{ij}^l = \pm 1$ 
for activating and inhibiting regulation, respectively, and 
$c_{ij}^l = 0$ if gene i does not receive an input from gene 
$j$ in cell $l$. The dynamics then is defined as
\begin{equation}
\label{signfunc}
\sigma_{j}^i(t) = \mbox{sign}\,(f_j(t-1))
\end{equation}
with 
\begin{equation}
\label{thresholdfunc}
f_j(t)  = \sum_{k = 1}^2\sum_{l = i-1}^{i+1} c_{kj}^{l}\sigma_k^l + h_j 
\end{equation}
for the ``hidden'' genes (compare Fig. \ref{thresscheme})
, where $\sigma_k^l, k \in \{1,2\}$ is the state of the $k$th
pattern gene in cell $l$ (there are no couplings between the 
genes in the ``hidden'' layer). The threshold $h_j$ is given by
\begin{equation}
\label{hdef}
h_j = \sum_{k = 1}^2\sum_{l = i-1}^{i+1} |c_{kj}^l|-2,
\end{equation}
which implements a logical OR operation. For the ``output''
(pattern) genes $G_1$ and $G_2$ in cell $i$, one simply has 
\begin{equation}
\label{patfuncdef}
f_k(t)  = \sum_{l = 1}^{k_{in}^k} \sigma_{l} - k_{in}^k,
\end{equation}
i.e.\ the weights are all set to one, and the (negative) 
threshold equals the number of inputs $k_{in}^k$ that 
gene $G_k, k \in \{1,2 \}$ receives from the hidden layer 
genes (logical AND).

\section{Results for deterministic dynamics}

Let us briefly summarize the dynamics of the simple stochastic cellular 
automata model of spatial pattern formation based on local information transfer
\citep{Rohlf2005}, and its de novo pattern formation by generating 
and regulating a domain boundary. Subsequently, we will discuss in detail how 
this very general mechanism can emerge as a result of interacting nodes in coupled 
identical networks, as a model for gene regulation networks in interacting cells. 

\subsection{Cellular Automata Model}

The first major outcome of the cellular automata model is that  a number of $n = 3$ different states is
necessary and sufficient for this class of systems to solve the given pattern
formation task
\footnote{The case $n=2$ corresponds to the class of elementary (Wolfram)
cellular automata with a very restricted set of 256 possible update rules.
In our extensive genetic algorithm runs, no solution for the here considered problem
was found for $n = 2$, even under diverse variations of the boundary conditions}.
The update table of the fittest solution 
found during optimization runs, which solves the problem independent of 
system size for about 98 percent of (randomly chosen) initial conditions 
(i.e. has fitness $\Phi = 0.98$), is shown in Table I.

\begin{figure}[htb]
\begin{center}
\resizebox{85mm}{!}{\includegraphics{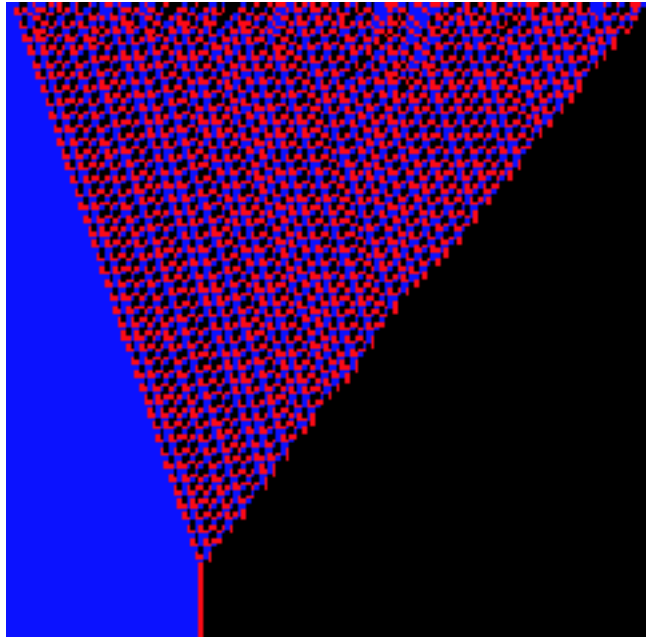}}
\end{center}
\caption{\small A typical dynamical run for the automata as defined in Table I, here for a system of size $N_C = 250$
cells (deterministic dynamics, no noise), starting from a random initial configuration. Time is
running on the $y$-axis from top to bottom. Cells with state $\sigma_i = 0$ are depicted in black color, cells with
$\sigma_i = 1$ in red and cells with $\sigma_i = 2$ in blue.  }
\label{ca1g}
\end{figure}

\begin{figure}[htb]
\begin{center}
\resizebox{85mm}{!}{\includegraphics{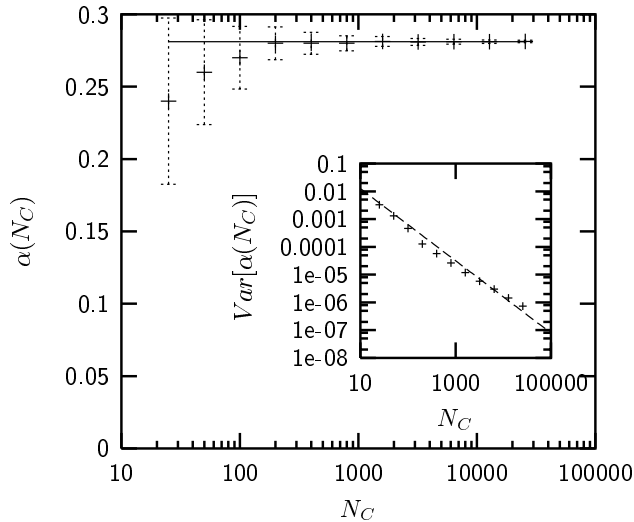}}
\end{center}
\caption{\small Finite size scaling of the self-organized relative domain size
$\alpha$ as a function of the total number of cells $N_C$. In the limit of large
system sizes, $\alpha$ converges towards a fixed value $\alpha_{\infty} = 0.281 \pm 0.001$
(as denoted by the straight line fit). The inset shows the finite size scaling
of the variance $Var[\alpha(N_C)]$; the straight line in this log-log plot has slope
$-1.3$ and indicates that fluctuations vanish with a power of the system size.}
\label{ndep}
\end{figure}

\begin{table}
\begin{center}
\begin{tabular}{|c|ccc|c||c|ccc|c|}\hline
index & $\sigma_{i-1}$ & $\sigma_{i}$ & $\sigma_{i+1}$ & $\sigma_i$ & 
index & $\sigma_{i-1}$ & $\sigma_{i}$ & $\sigma_{i+1}$ & $\sigma_i$ \\ \hline
0 & 0 & 0 & 0 & 0 & 14 & 1 & 1 & 2 & 2\\ \hline
1 & 0 & 0 & 1 & 2 & 15 & 1 & 2 & 0 & 0 \\ \hline
2 & 0 & 0 & 2 & 1 & 16 & 1 & 2 & 1 & 0\\ \hline
3 & 0 & 1 & 0 & 0 & 17 & 1 & 2 & 2 & 1\\ \hline
4 & 0 & 1 & 1 & 2 & 18 & 2 & 0 & 0 & 0\\ \hline
5 & 0 & 1 & 2 & 2 & 19 & 2 & 0 & 1 & 0\\ \hline
6 & 0 & 2 & 0 & 1 & 20 & 2 & 0 & 2 & 0\\ \hline
7 & 0 & 2 & 1 & 2 & 21 & 2 & 1 & 0 & 1\\ \hline
8 & 0 & 2 & 2 & 2 & 22 & 2 & 1 & 1 & 2\\ \hline
9 & 1 & 0 & 0 & 0 & 23 & 2 & 1 & 2 & 2\\ \hline
10 & 1 & 0 & 1 & 1 & 24 & 2 & 2 & 0 & 1\\ \hline
11 & 1 & 0 & 2 & 1 & 25 & 2 & 2 & 1 & 2\\ \hline
12 & 1 & 1 & 0 & 0 & 26 & 2 & 2 & 2 & 2\\ \hline
13 & 1 & 1 & 1 & 1 &&&&&\\ \hline
\end{tabular}
\end{center}
\caption{\small Rule table of most successful cellular automata solution found during genetic algorithm search. In the left column,
the rule table index is shown, running from 0 to 26, in the middle column the three input
states at time $t$ are shown, the right column shows the corresponding output states at time $t+1$.}
\end{table}

Fig. \ref{ca1g} shows the typical update dynamics of this solution. 
The finite size scaling of the self-organized relative domain size
$\alpha$ as a function of the number of cells $N_C$ is shown in Fig. \ref{ndep}. In the limit of large
system sizes, $\alpha$ converges towards  
\begin{equation}\label{alphavalue}
\alpha_{\infty} = 0.281 \pm 0.001.
\end{equation}
The variance of $\alpha$ vanishes with a power of $N_C$, i.e.\ the relative
size of fluctuations induced by different initial conditions becomes
arbitrarily small with increasing system size. Hence, the pattern
self-organization in this system exhibits considerable robustness against
fluctuations in the initial conditions. The main mechanism leading
to stabilization at $\alpha_{\infty} = 0.281$ is a modulation of
the traveling velocity $v_r$ of the right phase boundary in Fig. \ref{ca1g}
such that the boundary on average moves slightly
less than one cell to the left per update step, whereas the left boundary
moves one cell to the right exactly every third update step ($v_l = 1/3$). 
The modulation of the right boundary can be seen as the result of interacting phase boundaries 
reminiscent of particle interactions. This picture of 
``particle computation'' is a useful concept also in various 
other contexts \citep{Crutchfield1995}. 

From the fact that cells interact only with nearest neighbors one might conclude
that three cells in a row in principle would be sufficient to generate a pattern, which would be
in clear contradiction with a substantially larger minimum size of aggregates that was found, for example,
in the case of Hydra \citep{Technau2000}. The results summarized in Fig. \ref{ndep}, however,
indicate that pattern formation becomes very inprecise for systems smaller than $100$ cells and typically
fails for $N < 25$. Hence, the proposed mechanism is compatible with a required minimum
system size that substantially exceeds the range of local communication. While in reaction-diffusion
based models of pattern formation a certain extension of the field is required in order that the different
diffusion rates come into play, in our model the differential propagation of phase boundaries leads to a similar
effect.

\begin{figure}[htb]
\begin{center}
\resizebox{85mm}{!}{\includegraphics{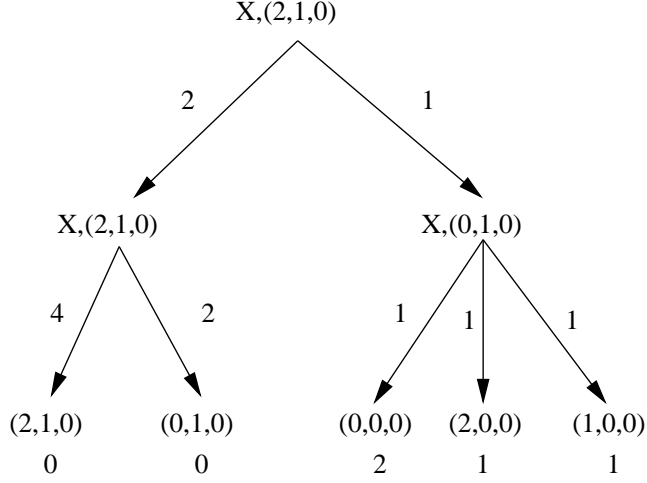}}
\end{center}
\caption{\small Transition tree for the boundary configuration $(2,1,0)$. Depending on the state of the left cell
$X$, transition to different configurations occur. Numbers on arrows indicate the total number of the respective
branches, the numbers at the bottom are the velocity (boundary readjustment) contributions of the respective
branches. For details, see text.}
\label{tree1}
\end{figure}

\begin{table}
\begin{center}
\begin{tabular}{|c|c|c|}\hline
configuration & $p_{ij0}$ & $\langle v\rangle_{ij0}$ \\ \hline
$(2,1,0)$ & 0.1646 & 0.257  \\ \hline
$(0,2,0)$ & 0.1852 & 0.591  \\ \hline
$(2,2,0)$ & 0.2058 & 0.167  \\ \hline
$(1,2,0)$ & 0.1646 & 0.862  \\ \hline
$(1,1,0)$ & 0.1317 & 0.778   \\ \hline
$(0,1,0)$ & 0.1482 & 2.629  \\ \hline

\end{tabular}
\end{center}
\caption{\small The six possible configurations at the right phase boundary
 with their respective probabilities $p_{ij0}$ and velocity contributions $\langle v\rangle_{ij0}$
of the corresponding transition trees (compare Fig. \ref{tree1}).}
\label{treetable}
\end{table}

Let us now derive a quantitative model that approximates the system dynamics. 
Since the left phase boundary travels at a constant speed of $v_l = 1/3$, we only have to derive
a (stochastic) model for the absolute value  $|\langle v_r\rangle|$
of the average traveling speed of the right phase boundary; the
equilibrium boundary position then follows as
\begin{equation}\label{alphaonv}
\alpha = \frac{v_l}{|\langle v_r\rangle| + v_l}.
\end{equation}
At the right phase boundary, there are three configurations (local update neighborhoods) that do not
lead to a readjustment of the boundary (namely, $(0,2,0)$, $(2,1,0)$ and $(2,2,0)$, the zero on the right marks the boundary). 
The configurations $(1,2,0)$ and $(1,1,0)$ readjust the boundary one cell to
the left, whereas the (extended) configurations $(x_1, x_2,0,1,0)$ move the boundary either two or three
cells leftward, depending on the states $x_1$ and $x_2$. Using a Markovian approximation (i.e., a one-step master
equation neglecting transition correlations between the six boundary configurations), $|\langle v_r\rangle|$ is approximated by
\begin{equation}\label{markovca1}
|\langle v_r\rangle| = 0\cdot p_0 + 1\cdot p_1 + 2\cdot p_2 + 3\cdot p_3,
\end{equation}
where $p_i$ are the respective probabilities to have a configuration that leads to boundary readjustment $i$ cells
at the left at the next time step. 
We neglect the slight asymmetries in the rule table and assume that each state $\sigma \in \{0,1,2\}$ appears with
probability $1/3$, hence it is straight-forward to derive $p_0 = 1/2$ and $p_1 = 1/3$. A slightly more detailed
analysis yields $p_2 = 1/18$ and $p_3 = 1/9$, leading to 
\begin{equation}\label{vr1}
|\langle v_r \rangle_{1}| \approx 0.78,
\end{equation}
which is about $9 \%$ below the true value $|\langle v_r \rangle| \approx 0.852$ measured in model simulations.

To improve the approximation, we now take into account transition correlations between different configurations and the slightly
asymmetric state distribution in the rule table. For each of the six boundary configurations,
a transition tree similar to Fig. \ref{tree1} is derived (for the last configuration, this tree consists of only one time
step and one transition, i.e. collapses on the Markovian approximation). Taking the average over 
the velocity contributions $v_{2i}$ of all branches of the
second time step (where $i$ numbers the branches), the contribution of the whole tree to 
the average phase boundary velocity per time step is
\begin{equation}\label{vtree}
 \langle v\rangle_{tree} = \frac{1}{2\,n_2}\sum_{i=1}^{n_2}v_{2i},
\end{equation}
where $n_2$ is the number of branches (here, $n_2 = 9$).  
The start configurations with their respective probabilities $p_{ij0}$ and velocity contributions $\langle v\rangle_{ij0}$
of the corresponding transition trees are listed in table 1. The phase boundary velocity now is calculated as
the weighted average
 \begin{equation}\label{markovca2}
|\langle v_r\rangle| =\sum_{i=0,j=1}^{2}p_{ij0}\cdot |\langle v\rangle_{ij0}|.
\end{equation}
Inserting the values of table \ref{treetable}, one finds
\begin{equation}\label{vr2}
|\langle v_r \rangle_{2}| \approx 0.82.
\end{equation}
Obviously, this value is a much better estimate than the zero-order approximation $\langle v_r \rangle_{1}$,
but still $4\%$ below the value $|\langle v_r \rangle| \approx 0.852$ 
measured in simulations. We conclude that this difference is an effect
of higher order correlations not included in our analysis.

\subsection{Interaction topology of the minimal network}
In this section, let us derive the structure of a minimal Boolean
network that solves the pattern formation problem, based on the previously discussed
results for cellular automata. We will see that this network
has biologically realistic properties regarding the number
of genes necessary for information processing and the complexity
of interaction structure, making it well conceivable that
similar ``developmental modules'' exist in biological systems.

\subsubsection{Boolean network model}

\begin{figure}[htb]
\begin{center}
\resizebox{85mm}{!}{\includegraphics{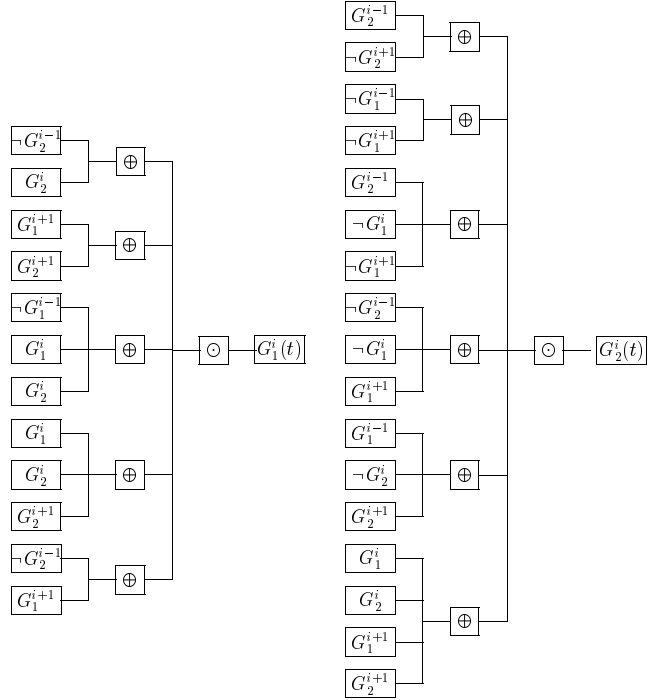}}
\end{center}
\caption{\small Boolean representation of the minimal network,
minimized conjunctive normal form. $G_a^b$
with $a \in \{1,2\}$ and $b \in \{i-1,i,i+1\}$
denotes gene $a$ in cell number $b$. The inputs in the
left branches of the trees are given by the genes'
states at time $t-1$. $\neg$ denotes NOT, $\odot$ denotes logical AND and $\oplus$ logical OR.   }
\label{logBoolean}
\end{figure}

\begin{figure}[htb]
\begin{center}
\resizebox{85mm}{!}{\includegraphics{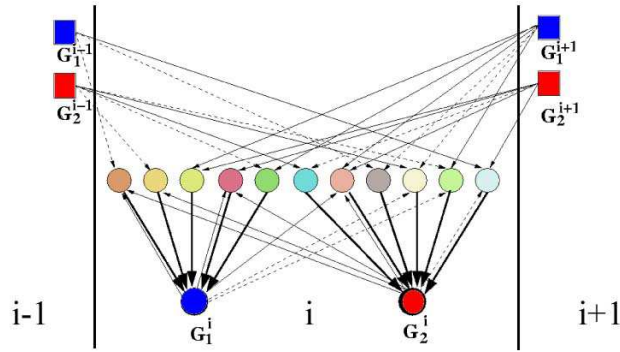}}
\end{center}
\caption{\small   Threshold network realization of the pattern formation system. Solid
line arrows denote links with $w_{ij} = +1$, dashed arrows denote links with $w_{ij} = -1$.
The inputs from the genes in the neighbor cells ($G_1^{i-1}$,$G_2^{i-1}$,  $G_1^{i+1}$ and  $G_1^{i+1}$)
are processed by a layer of ``hidden genes'' (colored circles in the middle of
the scheme) with different thresholds $h$ implementing a logical OR operation on the inputs.
The processed signals then are propagated to the two pattern genes $G_1$ and $G_2$. 
The threshold of gene $G_1$ is $h_1 = -5$, for gene $G_2$ one has $h_2 = -6$ (logical AND).
Notice several feed-back connections from genes $G_1^i$ and $G_2^i$ to the hidden layer.}
\label{threshold1}
\end{figure}

\begin{figure}[htb]
\begin{center}
\resizebox{60mm}{!}{\includegraphics{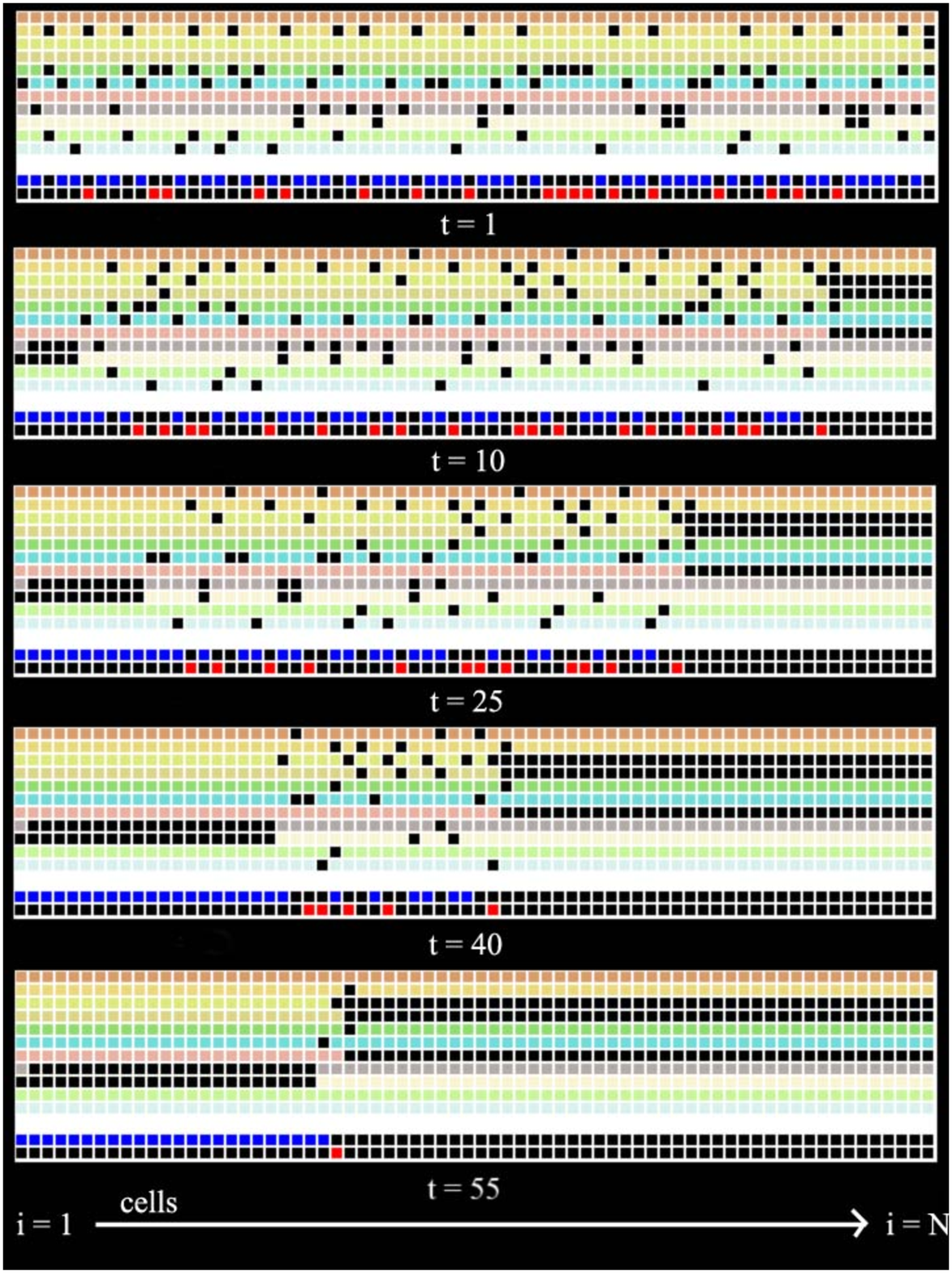}}
\end{center}
\caption{\small Snapshots of spatial gene activity patterns of the network shown in Fig. 6., for
a system of 80 coupled networks (cells) at different update times $t$. 
Each row shows the state of one network gene, color coding is
the same as in Fig. \ref{threshold1}; if the respective gene is not active in the cell at time $t$, the cell is shown
in black color. The two bottom rows show the states of the two target (pattern) genes.
After 55 update time steps, the target pattern (compare Fig. \ref{ca1g}) has self-organized.}
\label{genepat1}
\end{figure}

Let us now undertake the first step from the previous, coarse-grained
model of pattern formation to a detailed model that takes
into account the information processing capacity of cell-internal
regulatory networks, that can communicate locally with 
neighboring cells.
The rule table summarized in Table 1 is easily formalized in 
binary coding, i.e $0 \rightarrow 00$, $1 \rightarrow 01$
and $2 \rightarrow 10$, this corresponds to two ``genes'' 
$G_1$ and $G_2$ one of which ($G_1$) is active only in 
a domain at the left side of the cell chain. 
The so obtained Boolean update table is reduced to its minimized
conjunctive normal form, using a Quine-McCluskey
algorithm \citep{McCluskey1956}. For the construction of the network topology we use
the conjunctive normal form, as it is a somewhat biologically
plausible solution with a minimal number of logical AND operations.
In principle, other network topologies, e.g.\ with more levels of
hierarchy, are possible and biologically plausible, however, 
they involve a higher number of logical sub-processing steps, i.e.\ a
higher number of genes, hence we will not discuss them here.

Considering the huge number of possible input configurations which the outputs
theoretically could depend on, the complexity of the resulting network is surprisingly
low. As shown in Fig.\ \ref{logBoolean}, the output state of gene $G_1$ only depends
on five different input configurations of at maximum four different inputs,
gene number two on six different input configurations of at maximum four different inputs.
This indicates that the spatial information flowing into that network is strongly
reduced by internal information processing (only a small number of input states leads
to output ``1''), as expected for the simple stationary target pattern. Nevertheless,
this information processing is sufficient to solve the non-trivial task of domain
size scaling.

\subsubsection{Coupled threshold network model}

Threshold dependence of the states of regulatory elements
constitutes a biochemically simpler paradigm of switching behavior; information
processing dynamics is encoded in activating and inhibiting interactions only,
without the need for complex Boolean update tables. The simpler switching dynamics
comes at the expense of an increased network size, hence the formalization
as a threshold network gives us an estimate for the upper limit of regulatory network size
needed to solve the pattern formation problem. The coupled threshold network system, that
was derived according to the method outlined in section \ref{tn_method_subsec}, is shown in Fig.\ \ref{threshold1}. 
 The states of the genes $G_1$ and $G_2$
at time $t$ in a cell $i$ and its two neighbor cells $i-1$ and $i+1$
serve as inputs of 11 information-processing genes (``hidden'' layer). The state of these
genes then defines the state of $G_1$ and $G_2$ in cell $i$ at time $t+2$ (output layer).
Additionally, there is some feedback from $G_1$ and $G_2$ to the information processing layer,
as expected for the dependence on cell-internal dynamics already present in the cellular automata 
implementation of the model.

The resulting stationary spatial patterns of 
the information-processing 'hidden' genes 
and of genes $G_1$ (the ``domain gene'') and $G_2$
(active only at the domain boundary)   are shown in Fig. \ref{genepat1} (snapshots of five different
update time steps for a system of 70 cells). 
Starting from a random initialization of the two pattern genes $G_1$ and $G_2$,
due to the high-level genetic information processing in the hidden layer the target pattern
self-organizes robustly within 55 update time steps.
The network we construct here, regarded as a ``developmental module'' defining the
head-foot polarity through spatially asymmetric gene expression, has a size similar to 
comparable biological modules (compare, for example, the segment polarity network in \emph{Drosophila} \citep{Dassow2000,Albert2003})
as well as similar complexity (e.g., average connectivity $\bar{K} \approx 3$).

\begin{figure}[htb]
\begin{center}
\resizebox{85mm}{!}{\includegraphics{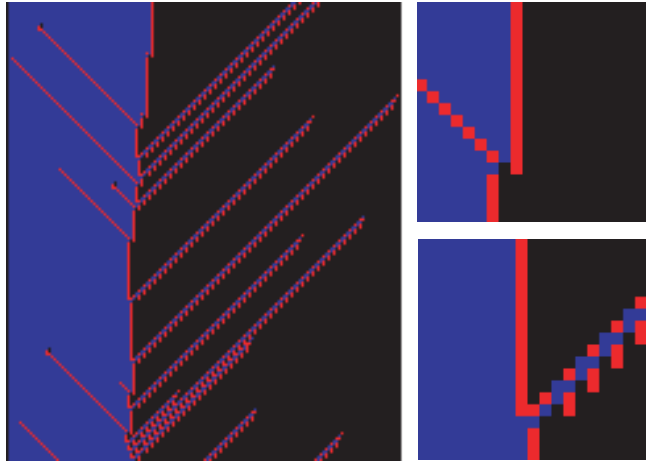}}
\end{center}
\caption{\small Quasi-particles, started by stochastic update errors, lead to control of the boundary
position under noise (left panel). The $\Gamma$ particle (top right) leads to readjustment of the boundary two cells to the left,
the $\Delta$ particle (bottom right) leads to readjustment of the boundary one cell to the right. }
\label{particles}
\end{figure}

\begin{figure}[htb]
\begin{center}
\resizebox{85mm}{!}{\includegraphics{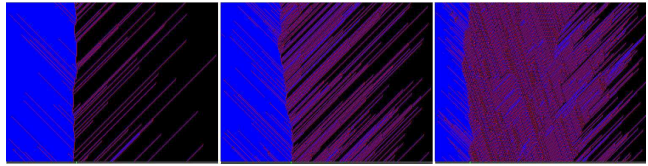}}
\end{center}
\caption{\small For moderate error rates $r_e$, the domain boundary is stabilized at an average position $\alpha^* =
1/3$ (left panel, $r_e = 0.1$). Around $r_e \approx 0.2$, there is a crossover to a domain size vanishing with $r_e^{-1}$ (middle panel). In the right panel, the high noise limit is shown, with a considerably shrinked blue domain due to
strong particle interference ($r_e = 2.0$).}
\label{cacompnoi}
\end{figure}

\begin{figure}[htb]
\begin{center}
\resizebox{85mm}{!}{\includegraphics{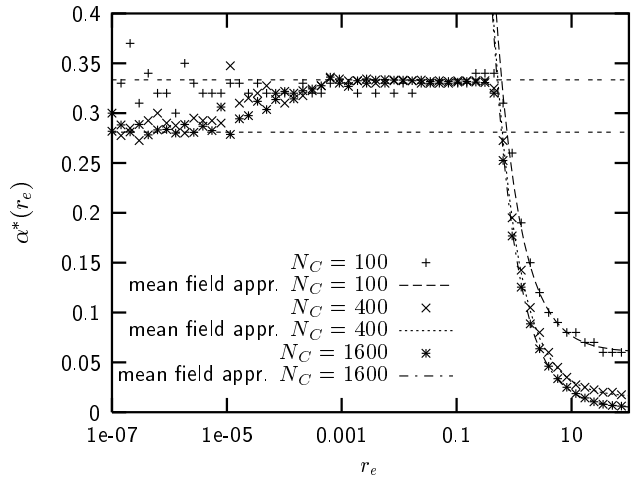}}
\end{center}
\caption{\small Average boundary position $\alpha^*$ as a function of the error rate $r_e$ for system sizes
$N_C = 100$, $N_C = 400$, and $N_C = 1600$. The abscissa is logarithmic. Numerical data are averaged over $200$ different initial conditions with
$2\cdot10^6$ updates each. The dashed curves show the mean field approximation given by Eqn.\ (\ref{rate}), the straight dashed lines
mark the unperturbed solution $\alpha^* = 0.281$ and the solution under noise, $\alpha^* = 1/3$. }
\label{phase}
\end{figure}

\begin{figure}[tb]
\begin{center}
\resizebox{85mm}{!}{\includegraphics{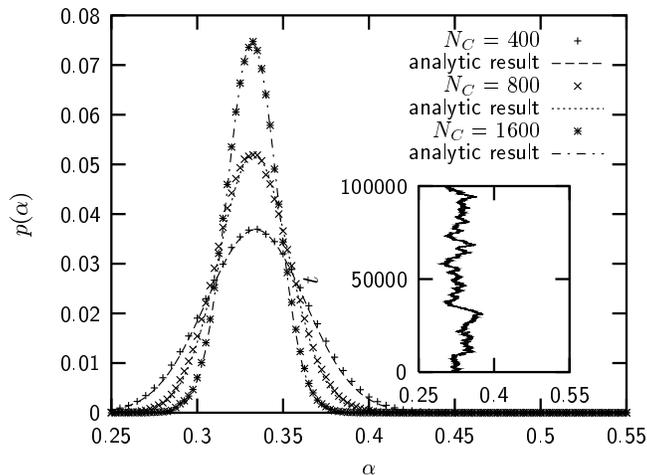}}
\end{center}
\caption{\small For the system with stochastic update errors, fluctuations of the boundary position $\alpha$
around the average position $\alpha^* = 1/3$ are Gaussian distributed. The figure compares the numerically
obtained stationary probability distribution with the analytic result of Eqn.\ (\ref{statsol2}) for three different system sizes. All data are gained for
$r_e = 0.1$ and averaged over $100$ different initial conditions with $2\cdot 10^6$ updates each. The inset shows a typical timeseries
of the boundary position. }
\label{fluct}
\end{figure}
\begin{figure}[htb]
\begin{center}
\resizebox{85mm}{!}{\includegraphics{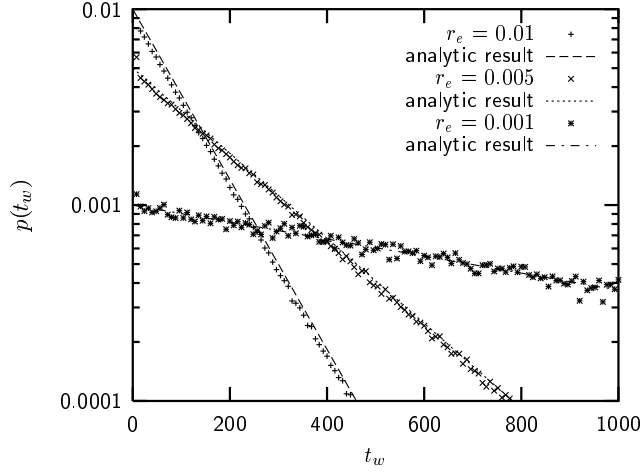}}
\end{center}
\caption{\small Probability distribution $p(t)$ of waiting times for boundary readjustments in the model
with stochastic update errors for three different values of $r_e$, semi-log plot. As expected for a Poisson process,
$p(t)$ is an exponential. }
\label{wait}
\end{figure} 

\begin{figure}[htb]
\begin{center}
\resizebox{85mm}{!}{\includegraphics{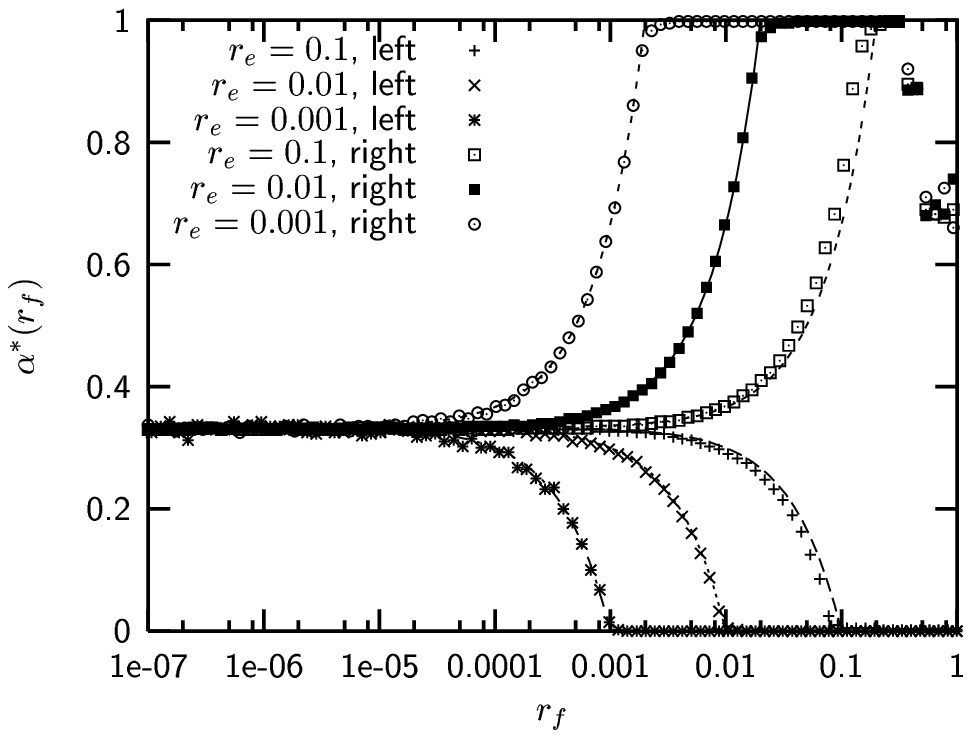}}
\end{center}
\caption{\small Average domain size $\alpha^*$ as a function of the the cell flow rate $r_f$ for three
different error rates $r_e$; numerical data (crosses and points) were sampled over 10 different initial conditions and 1e6
updates for each data point. ``Left'' indicates cell flow directed to the left system boundary, ``right'' to the right system boundary,
respectively. The dashed lines are the corresponding solutions of Eqn.\ (\ref{20}).}
\label{cnflow}
\end{figure}

\section{Dynamics under noise and cell flow}\label{noise_sec}

In the following we will study dynamics and robustness of the model with respect to noise. 
Two kinds of perturbations frequently occur: Stochastic update errors and external
forces induced by a directed \emph{cell flow} due to cell proliferations. Both types
of perturbations are very common during animal development, e.g., in  \emph{Hydra} cells continuously move from the central 
body region along the body axis towards the top and bottom, and differentiate into the respective
cell types along the way according to their position on the head-foot axis. 
However, similar problems of reliable pattern formation in noisy and 
highly dynamical environments occur in tissues with a high turnover. 
High proliferation rates and/or movement of cells occur, for example, 
in skin tissues and when stochastic phenomena of cell (re-)differentiation, 
e.g. stochastic stem cell production, are found. 
In our model, we abstract stochastic changes in differentiation
states by stochastic update errors, and consider directed cell movements
in the form of a steady cell flow.

Let us define stochastic update errors with probability $p$ per cell,
leading to an average error rate $r_e = p\,N_C$. Interestingly, 
this stochastic noise starts moving ``particle'' excitations in the cellular 
automaton which, as a result indeed stabilize the developmental structure of 
the system. To prepare for the details of these effects, define first how we 
measure the boundary position properly in the presence of noise.  
Let us use a statistical method to measure the boundary position
in order to get conclusive results also for high $p$: Starting at $i=0$,
we put a ``measuring frame'' of size $w$ over cell $i$ and the next $w-1$ cells,
move this frame to the right and, for each $i$,  measure
the fraction $z$ of cells with state $\sigma = 2$ within the frame. The algorithm
stops when $z$ drops below $1/2$ and the boundary position is defined to be $i + w/2$.

One can show that, for not too high $p$, there are only
two different quasi-particles (i.e. state perturbations
moving through the homogeneous phases), as shown in Fig. \ref{particles}. In the following,
these particles are called $\Gamma$ and $\Delta$. The  $\Gamma$
particle is started in the $\sigma_2$ phase by a stochastic error $\sigma_i = 2
 \rightarrow \sigma_i \ne 2$ at some $i < \alpha N_C$), moves to the right and,
when reaching the domain boundary, readjusts it
two cells to the left of its original position. The $\Delta$ particle is started in the $\sigma_0$
phase by a stochastic error $\sigma_i = 0 \rightarrow \sigma_i \ne 0$ at
some $i > \alpha N_C$ and moves to the left. Interaction with the domain
boundary readjusts it one cell to the right. Thus we find that the average position $\alpha^*$ of the
boundary is given by the rate equation
\begin{equation}
 2\alpha^* r_e = (1 - \alpha^*) r_e,
\label{rate}
\end{equation}
i.e. $\alpha^* = 1/3$. Interestingly, for not too high error rates $r_e$,  $\alpha^*$
is independent from $r_e$ and thus from $p$. If we consider the average boundary position  $\alpha^*$ as
a system-specific \emph{order parameter} which is controlled by the two quasi-particles,
then comparing the solution of Eqn.\ (\ref{rate}) to the equilibrium position in the noiseless case indicates
that the system undergoes a step-like discontinuity with respect to  $\alpha^*$ at $p = r_e = 0$.
This conclusion is supported by a numerical analysis of the finite size scaling of this transition
(cf. appendix \ref{noitrans_app}).

The solution  $\alpha^* = 1/3$ is stable only for $0 < r_e \le 1/2$. 
As shown in Fig.\ \ref{particles}, the interaction of a  $\Gamma$ particle with the boundary
needs only one update time step, whereas the boundary readjustment following a $\Delta$ particle 
interaction takes three update time steps. Hence, we conclude that the term on the right hand side of Eqn.\ (\ref{rate}), 
which gives the flow rate of $\Delta$ particles at the boundary, for large $r_e$ will saturate at $1/3$, leading to
\begin{equation} \label{alphasat}
2\alpha^* r_e = \frac{1}{3} 
\end{equation} 
with the solution
\begin{equation} \alpha^*  = \frac{1}{6}\,\, r_e^{-1} + \Theta(N_C) \label{alphasatsol}\end{equation}
for $r_e > 1/2$. Hence, there is a crossover from the solution $\alpha^* = 1/3$ to another solution vanishing
with $r_e^{-1}$ around  $r_e = 1/2$.
The finite size scaling term $\Theta(N_C)$ can be estimated from the following
consideration: for $p \rightarrow 1$, the average domain size created by ``pure chance''
is given by $\alpha^* = N_C^{-1} \,\sum_{n=0}^{N_C} (1/3)^n \cdot n \approx (3/4)\,N_C^{-1}$. If the measuring
window has size $w$, we obtain $\Theta(N_C) \approx (3/4)\,w\,N_C^{-1}$.
To summarize, we find that the self-organized boundary position is given by
\begin{equation}\label{allnoisesol}
\alpha^*=  \left\{  
\begin{array}{cccc} 0.281 \pm 0.001  \quad &\mbox{if}& r_e =0 \\
 1/3  \quad &\mbox{if}& 0 < r_e \le 1/2 \\
         (1/6)r_e^{-1} + \Theta(N_C)  \quad &\mbox{if}&  \quad r_e > 1/2 &
\end{array} \right. 
\end{equation}
with a step-like discontinuity at $r_e = 0$ and a crossover around $r_e = 1/2$. 

Now let us consider the fluctuations of $\alpha$ around $\alpha^*$ given by the master equation
\begin{eqnarray} p^{\tau}(\alpha) &=& 2\alpha\,r_e\, p^{\tau-1}(\alpha + 2\delta) + (1-\alpha)\,r_e\, p^{\tau-1}(\alpha - \delta) \nonumber  \\
&+& (N_C-r_e)\, p^{\tau-1}(\alpha) - 2\alpha\,r_e\, p^{\tau-1}(\alpha) \nonumber \\  &-& (1-\alpha)\,r_e\, p^{\tau-1}(\alpha)\label{master1}
\end{eqnarray} 
with $\delta = 1/N_C$. Eqn.\ (\ref{master1})
 determines the probability $p^{\tau}(\alpha)$ to find the boundary at position $\alpha$
at update time step $\tau$, given its position at time $\tau-1$. This equation can be simplified as we are interested only
in the \emph{stationary probability distribution} of $\alpha$. It is easy to see that the  error rate $r_e$ just provides 
a time scale for relaxation towards
the stationary distribution and has no effect on the stationary distribution itself. Therefore, we may consider the limit $r_e \to r_e^{max} := N_C$,
divide through $r_e$ and neglect the last three terms on the right handside of Eqn.\ (\ref{master1}) (which become
zero in this limit). We obtain 
\begin{equation} p^{\tau}(\alpha) = 2\alpha\, p^{\tau-1}(\alpha + 2\delta)  + (1-\alpha)\,  p^{\tau-1}(\alpha - \delta).
\label{master2}  \end{equation} 
To study this equation, we consider the continuum limit $N_C \to \infty$.
Let us introduce the scaling variables $x = (\alpha - \alpha^*)\sqrt{N_C}$, $t = \tau /N_C$ and the probability density
$f(x,t) = N_C\, p^{\tau}(\alpha\, N_C)$. Inserting these definitions into Eqn.\ (\ref{master2}) and ignoring all subdominant powers $\mathcal{O}(1/N_C)$,
we obtain a Fokker-Planck equation:
\begin{equation} \frac{\partial f(x,t)}{\partial t} = \left( \frac{\partial^2}{\partial x^2} + 3 \frac{\partial}{\partial x}\,x\right) f(x,t).
\label{fokkerplanck} \end{equation} 
The stationary solution of this equation is given by
\begin{equation} f(x) = \sqrt{\frac{3}{2\pi}}\exp{\left[-\frac{3}{2}x^2 \right]},
 \label{statsol1}\end{equation} 
i.e. in the long time limit $t \to \infty$, the probability density for the boundary position $\alpha$ is a Gaussian with mean $\alpha^*$:
\begin{equation} p(\alpha, N_C ) = \sqrt{\frac{3\,N_C}{2\pi}}\exp{\left[-\frac{3\,N_C}{2}\left(\alpha-\alpha^*\right)^2 \right]}.\label{statsol2}
 \end{equation}
From Eqn.\ (\ref{statsol2}) we see that the variance of $\alpha$ vanishes $\sim 1/N_C$ and the relative boundary position becomes sharp
in the limit of large system sizes. Fig. \ref{fluct} shows that this continuum approximation for $N_C \ge 400$ provides
very good correspondence with the numerically obtained probability distributions.

The stochastic nature of boundary stabilization under noise is also reflected
by the probability distribution of \emph{waiting times} $t$ for boundary readjustments
due to particle interactions: the particle production is a Poisson process with
the parameter $\lambda = r_e$ and the waiting time distribution is given by
\begin{equation} p_{wait}(t) =  r_e\exp{(-r_e\,t)} \end{equation}
with an average waiting time $\langle t \rangle = r_e^{-1}$. Fig.\ \ref{wait} shows the waiting time
distributions for different error rates $r_e$.

In a biological organism, a pattern has to be robust not only with respect
to dynamical noise, but also with respect, e.g., to ``mechanical'' perturbations. In \emph{Hydra}, e.g., 
there is a steady flow of cells directed towards the animal's head and foot, due to continued proliferation
of stem cells \citep{David1972}; the stationary pattern of gene activity
is maintained is spite of this cell flow. Let us now study the robustness of the model with respect to this type of perturbation.
Let us consider a constant cell flow with rate $r_f$, which is directed
towards the left or the right system boundary. In Eqn.\ (9), we now get an additional drift term $r_f$
on the left hand side:
\begin{equation} 2\alpha^*r_e \pm  r_f = r_e(1 - \alpha^*),  \label{20}\end{equation} 
with the solution
\begin{equation}
\alpha^*=  \left\{  \begin{array}{cccc} \frac{1}{3}\left ( 1 - \frac{r_f}{r_e}\right )  \quad &\mbox{if}&  r_e \ge r_f \\ 
                                          0     \quad  &\mbox{if}&  r_e < r_f &
\end{array} \right. 
\end{equation}
for the case of cell flow directed towards the left system boundary (plus sign in eqn. (\ref{20})).
One observes that $\alpha^*$ undergoes a second order phase transition at the critical value $r_e^{crit} = r_f$. 
Below $r_e^{crit}$, the domain size $\alpha^*$ vanishes, and above $r_e^{crit}$ it grows until it reaches
the value $\alpha^*_{max}=1/3$ of the system without cell flow. For cell flow directed
towards the right system boundary (minus sign in eqn. (\ref{20})), we obtain
\begin{equation}
\alpha^*=  \left\{  \begin{array}{cccc} 
\frac{1}{3}\left ( 1 + \frac{r_f}{r_e}\right )  \quad &\mbox{if}&  r_f \le 2\,r_e \\ 
                   1     \quad  &\mbox{if}&  r_f > 2\,r_e.
\end{array} \right. 
\end{equation}
In this case, the critical cell flow rate is given by $r_f = 2\, r_e$, for cell flow rates larger
than this value the $\sigma_2$-domain extends over the whole system, i.e.\ $\alpha^* = 1$. 

Fig.\ \ref{cnflow} compares the results of numerical simulations
with the mean field approximation of Eqn.\ (\ref{20}). In numerical simulations, cell flow is realized by application
of the translation operator $\Theta\,\sigma_i := \sigma_{i+1}$ to all cells with $0 \le i < N_C-1$ every $r_f^{-1}$ time steps
and leaving $\sigma_{N_C-1}$ unchanged. In case of cell flow directed to the right system boundary, in the limit
$r_f \to 1$ the boundary position $\alpha^*$ detected in numerical simulations deviates from the mean field prediction,
due to a boundary effect at the left system boundary (stochastic production of finite lifetime stationary oscillators, leading
to intermittent flows of $\Gamma$ particles through the system).

To summarize this part, we see that in the model stochastic errors in dynamical updates for $r_e > r_f$ indeed \emph{stabilize} the global pattern
against the mechanical stress of directed cell flow.

\section{Proportion regulation in a growing system}
So far, we assumed that the system size $N$ (the number of cells) is constant, which is a good approximation for an adult organism;
in a developing organism, however, proportion regulation has to work under the condition
of a steadily growing system size. Here, we study this problem for two simplified settings: first, for symmetric growth,
i.e., new cells are
added with probability $1/2$ on either side of the chain of cells, and the growth rate $r_g$ is constant on average; second,
for homogeneous cell proliferation with probability $p_d$ per cell, assuming that daughter cells inherit the state of the mother cell.

\begin{figure}[htb]
\begin{center}
\resizebox{85mm}{!}{\includegraphics{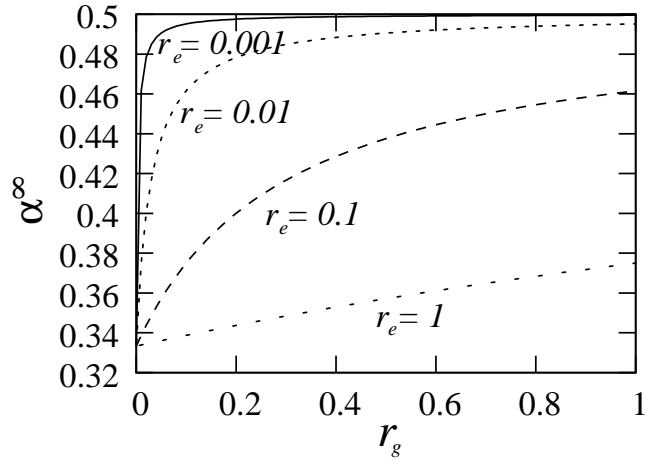}}
\end{center}
\caption{\small Asymptotic boundary position $\alpha^{\infty}(r_g)$ in the case of unlimited symmetric growth at
the boundaries of the cellular array, for four different error rates $r_e$. }
\label{growth}
\end{figure}
\begin{figure}[htb]
\begin{center}
\resizebox{85mm}{!}{\includegraphics{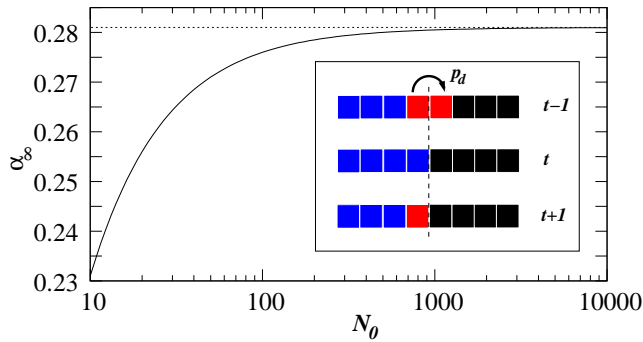}}
\end{center}
\caption{\small Asymptotic domain boundary position $\alpha_{\infty}$ (lined curve) as a function
of the initial system size $N_0$, for homogeneous growth, as explained in the text. One has 
$\alpha_{\infty} = \alpha_0 - 1/N_0$,
where $\alpha_0 = 0.281$ is the boundary position of the constant-size system (dashed line).
{\em Inset:} Proliferation of the boundary cell (red) at time $t-1$ leads to readjustment
of the boundary at its original position (indicated by the dashed line) at time $t+1$, thereby increasing the black domain
by one cell and hence slightly reducing $\alpha$.
}
\label{homgrowth_fig}
\end{figure}

\subsection{Symmetric growth at the system boundaries}
Let us assume we start with a system of $N_0$ cells, with an initial boundary position at cell $N_1$. In the deterministic case $r_e=0$, it is straight-forward to
see that the asymptotic boundary position in the limit of large times $t$ is given by
\begin{equation} 
 \alpha^{\infty} = \lim_{t\to\infty} \alpha^*(t) = \frac{1}{2}
 \end{equation}
(for details, see appendix \ref{symmgroapp}).
This means that in the limit $r_e = 0$, proportion regulation cannot be maintained under the condition of a steady system growth.
In the case $r_e > 0$ and assuming infinite growth, the asymptotic boundary position is given by
 \begin{equation} 
 \alpha^{\infty} = \lim_{t\to\infty} \alpha^*(t) = \frac{\frac{1}{2}r_g + r_e   }{r_g+3r_e}
 \end{equation}
(a derivation can be found in appendix \ref{symmgroapp}).
 Fig. \ref{growth} shows $\alpha^{\infty}(r_g)$ for four different values of $r_e$; it becomes evident that an approximately 'correct'
 proportion regulation requires $r_e$ to be at the order of $r_g$ or larger, i.e. $r_e/r_g \ge 1$. While $r_e$
 (the rate of regulatory signals) may not be increased significantly above the growth rate $r_g$, due to metabolic constraints,
 in later stages of development the steady decrease of $r_g$ will ensure that the condition $r_e/r_g \ge 1$ is fullfilled and proportion regulation
 approaches the steady state of the adult organism.
 
\subsection{Homogeneous cell proliferation}

Another simple case is system growth by {\em homogeneous cell proliferation.} 
Assuming that daughter cells inherit the state of their mother cell, one can show that proportion regulation
is maintained even in the case of zero noise, under the simplifying assumption that initial pattern formation takes place
in a system of size $N_0$, and that system growth does not start before pattern formation has converged to its attractor.
Due to an instability induced by proliferation events directly at the boundary cell with $\sigma_b = 1$ (compare inset
of Fig. \ref{homgrowth_fig}), slight deviations
asymptotic boundary position $\alpha_0$ of non-growing systems
are found for finite $N_0$ (Fig. \ref{homgrowth_fig}, for details, cf. appendix \ref{homgroapp}):

\begin{equation}
\alpha_{\infty} = \lim_{t\to\infty}\alpha(t)= \alpha_0 - \frac{1}{2N_0}
\end{equation}

For the case when noise is present, it is not possible to find a general solution since proliferation
can affect both the velocity and the type of particles travelling through the domains in intricate ways
(in the case of symmetric growth at the system boundaries,
as discussed in the previous subsection, this problem is avoided). If $p_d$ is very small,
however, we can assume that proliferation events and particle propagation are essentially decoupled
and that the system has enough time to relax to a stationary state between proliferation events. In this limit,
one can show that the asymptotic boundary position converges to the value $\alpha = 1/3$ of the
stationary size system as discussed in section \ref{noise_sec} (for a derivation cf. appendix \ref{homgroapp}).

\section{Regeneration in a simulated cut experiment}

\begin{figure}[htb]
\begin{center}
\resizebox{60mm}{!}{\includegraphics{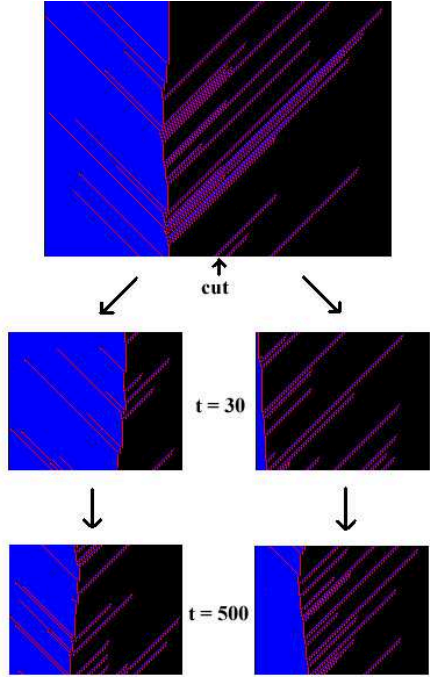}}
\end{center}
\caption{\small A simulated cut experiment. After $300$ updates, a system of $500$ cells was cut into two sub-systems of $250$ cells each (upper panel).
After only $30$ updates, in both sub-systems reorganization of the boundary position starts (middle panels). After about $500$ system updates, both sub-systems
have self-organized into the target pattern with $\alpha = 1/3$ (bottom panels). In the simulation, $r_e = 0.01$ was applied. }
\label{cutexp}
\end{figure}

Simple multi-cellular organisms as, e.g., {\em Hydra} exhibit remarkable regeneration capacities, which include, as already discussed, 
proportion regulation and de-novo pattern formation after complete dissociation of the body tissue. Similarly, it was already observed
in the late 19th century that polyps can be cut in half, leading to regeneration of two new, intact animals \citep{}. Without going
into the more intricate details of these experiments, we now demonstrate that, given minimum level of noise in the system, our model
in principle can reproduce this type of observation. Fig. \ref{cutexp} illustrates a simulated cut experiment, where, after 300 initial
system updates, the cellular array was cut into two equal-sized halves. After just 500 subsequent updates, both new sub-systems have
self-organized again into the target pattern with $\alpha = 1/3$.

\section{Robustness under noisy direction recognition}
While asymmetries in receptor distribution on cell membranes, asymmetric distribution of cell factors in the cell
or an extrenal gradient might provide some information about the asymmetry (the direction) of the
spatial pattern along the body axis, which then can be processed by a cell-internal gene regulatory network,
a substantial amount of noise can be expected to be present in this process. In particular, in the system
discussed in this paper, this type of information can be assessed only locally, hence we expect that local
errors in 'direction recognition' can substantially disrupt the emergence of the global pattern.

\begin{figure}[htb]
\begin{center}
\resizebox{85mm}{!}{\includegraphics{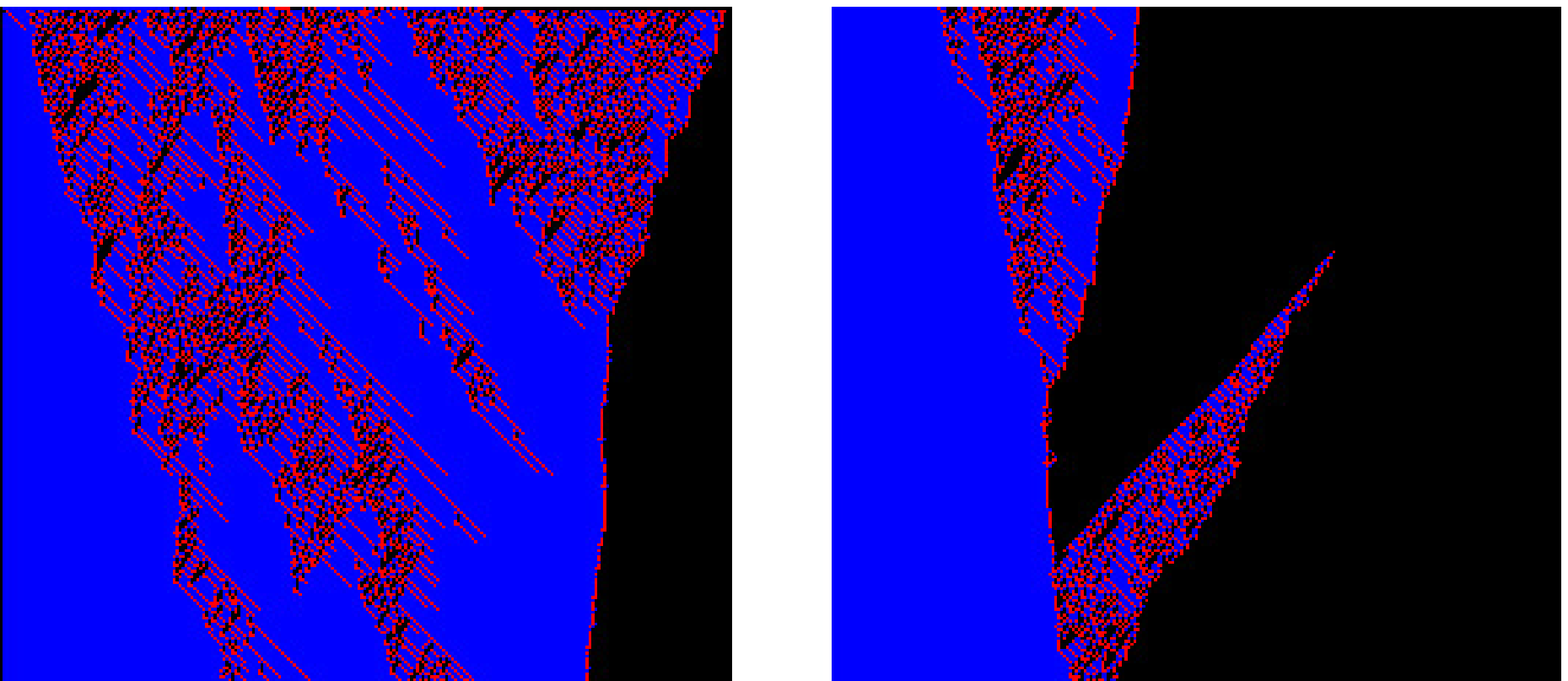}}
\end{center}
\caption{\small Dynamics of pattern formation under noisy direction recognition; parameters in the simulation shown here
were $p_{dir} = 0.1$ and $r_e = 0.003$. Initial pattern formation (left panel) as well as control
of the boundary by quasi-particles (right panel) still work, though particle trajectories become broadened and blurred.  }
\label{dirdyn_ca}
\end{figure}

\begin{figure}[htb]
\begin{center}
\resizebox{85mm}{!}{\includegraphics{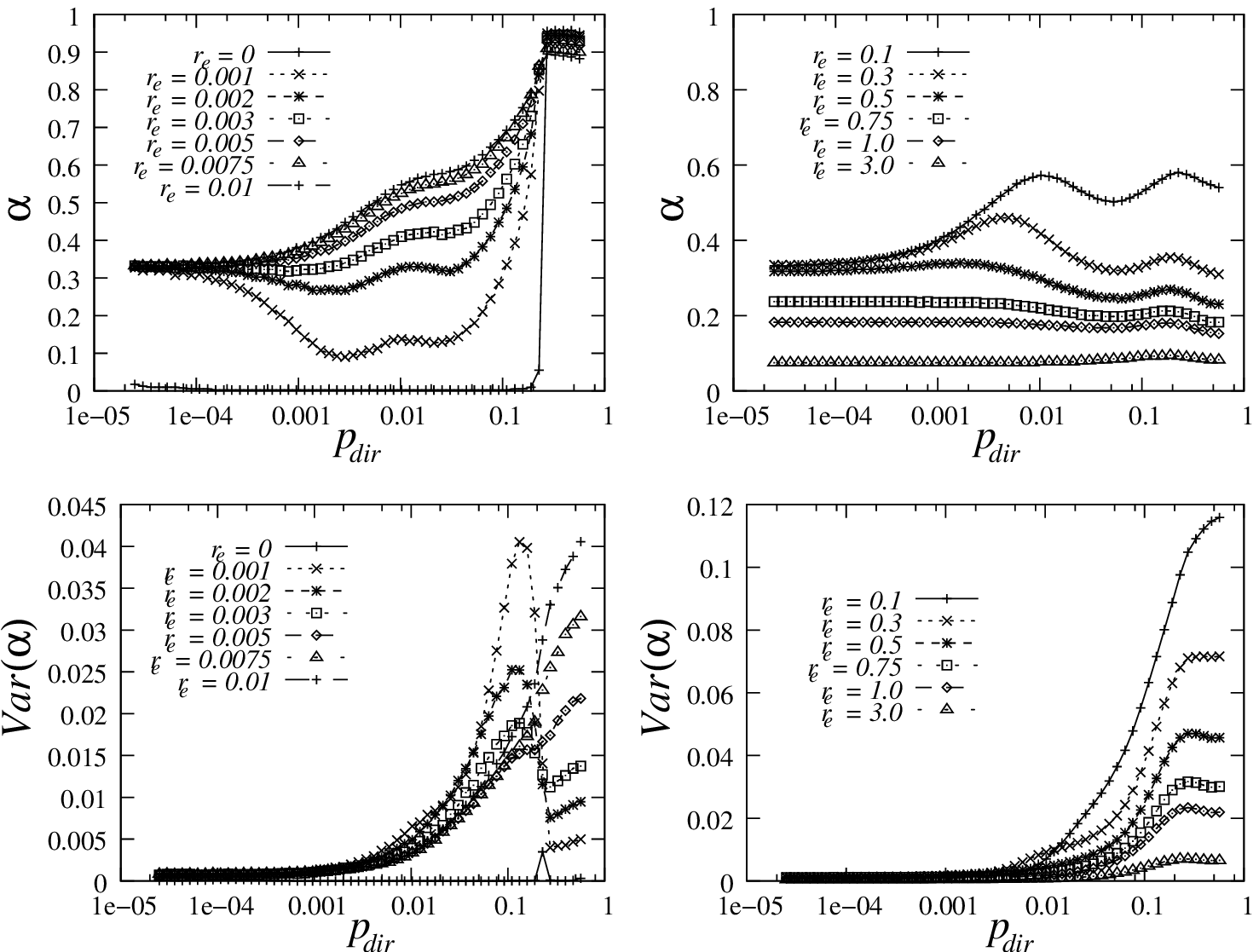}}
\end{center}
\caption{\small {\em Upper panels}: average boundary position $\alpha$ as a function of $p_{dir}$, for different values of $r_e$. Left panel: low
noise limit ($r_e \to 0$), right panel: high noise limit. {\em Lower panels}: Fluctuations (temporal variance) of the boundary position
as a function of $p_{dir}$. }
\label{dirdyn_quantitative}
\end{figure}

We now test the robustness of our model with respect to this type of errors. Let $(\sigma_{i-1}(t),\sigma_i(t),\sigma_{i+1}(t))$ be the state
of the neighborhood of cell $i$ at time $t$, then the state of cell $i$ at time $t+1$ is given by

\begin{equation}
\sigma_i(t+1) =  \left\{  \begin{array}{cccc} 
 f(\sigma_{i-1}(t),\sigma_i(t),\sigma_{i+1}(t)) &\mbox{with prob.}& 1 - p_{dir}\\ 
 f(\sigma_{i+1}(t),\sigma_i(t),\sigma_{i-1}(t)) &\mbox{with prob.}& p_{dir}
\end{array} \right.,
\end{equation}
where $p_{dir}$ is the probability of false direction recognition, and $f(.)$ is the corresponding rule table entry associated to the state 
 $(\sigma_{i-1}(t),\sigma_i(t),\sigma_{i+1}(t))$ and its locally inverted state $(\sigma_{i+1}(t),\sigma_i(t),\sigma_{i-1}(t))$, respectively.
 
Our first finding is that the dynamics of the original system (deterministic dynamics, i.e. $r_e = 0$), is indeed disrupted, due to a destabilization of the boundary
state. For $p_{dir} < 0.2$, $\alpha$ always goes to zero, while at $p_{dir} \approx 0.2$, there is an abrupt jump to $\alpha \approx 0.95$ (Fig. \ref{dirdyn_quantitative}, 
top left panel).
However, the situation changes substantially in the much more realistic case of a finite error rate in dynamical updates, i.e. $r_e > 0$. Fig. 
\ref{dirdyn_ca} demonstrates
that in this case initial pattern formation (left panel), as well as control of the boundary position by noise induced particles (right panel) work,
although the trajectory of the information-transmitting particles is blurred out and broadened by the stochastic errors in direction
recognition. The latter effect is reflected by an increase in fluctuation size (an increase of the variance) of the boundary position $\alpha$ with
increasing $p_{dir}$, in particular in the limit of high dynamical error rate $r_e$ (Fig. \ref{dirdyn_quantitative}, bottom panels). Remarkably, the {\em average} boundary
position $\alpha$ is stabilized over a wide range of the new control parameter $p_{dir}$, both in the limit of small $r_e$ (Fig. \ref{dirdyn_quantitative}, 
left upper panel) and large
$r_e$ (Fig. \ref{dirdyn_quantitative}, right upper panel). While there is some dependence on both $p_{dir}$ {\em and} $r_e$, as reflected by the fact that the curves $\alpha(p_{dir}, r_e)$ for
different values of $r_e$ do not collapse, the principal pattern formation mechanism still works, given we stay at reasonable values $p_{dir} < 0.2$. Hence, the
system exhibits remarkable robustness also with respect to errors in direction recognition. Notice that this robustness was {\em not selected for}
in GA runs, i.e., it is a truly emergent property of the system dynamics. In a real system of coupled gene regulatory networks,
additional mechanisms for error correction might be present, that, for example, process information not only from direct neighbor cells, or 
exploit the {\em 2D or 3D geometry} of a real tissue \citep{Rohlf2004b}. 

\section{Summary and Discussion}
In this paper we considered the dynamics of pattern formation motivated by animal morphogenesis 
and the largely observed participation of complex gene regulation networks in their coordination and control. 
We therefore chose a simple developmental problem to study toy models of interacting networks that 
control pattern formation and morphogenesis in a multicellular setting. In particular, the goal was to explore 
how networks can offer additional mechanisms beyond the standard diffusion based process of the 
Turing instability. Our results suggest that main functions of morphogenesis can be performed by 
dynamical networks without relying on diffusive biochemical signals, but using local signaling between 
neighboring cells. This includes solving the problem of generating global position information from purely
local interactions, but also it goes beyond diffusion based models as it offers solutions to developmental 
problems that are difficult for such models and avoids their inherent problem of fine tuned model parameters. 
Indeed, it has been shown in case studies that this paradigm applies well to development, as for example
for the segment polarity network of {\em Drosophila}, which exhibits robustness against
parameter variations by several orders of magnitude \citep{Dassow2000}, and where spatial
gene expression patterns can be predicted reliably from the topology of regulatory interactions
alone in a Boolean network model \citep{Albert2003}.
In many cases, developmental processes as, e.g., the establishment of positional information, may rely on this type of 
internal information processing rather than on interpretation of global chemical gradients. 
In this type of local information processing, several ways how spatial symmetry of morphogenetic
signals could be broken are conceivable. Cells potentially could exploit local anisotropies in receptor localization \citep{Galle2002},
as well as gradients produced by local propagation of morphogens \citep{Kasatkin2007} or juxtacrine signaling \citep{Monk1998}.
In either case, the mechanism proposed in our model would exhibit considerable robustness, since only
a rough estimate on the direction of the receptor anistropy or the gradient is needed (compare section 7
on robustness under noisy direction recognition. 

The network model derived here performs accurate regulation of position information and robust {\em de novo} pattern formation
from random conditions, with a mechanism based on local information transfer rather than the Turing instability. 
Non-local information is transmitted through soliton-like quasi-particles instead of long-range gradients.
Two realizations as discrete dynamical networks, Boolean networks and threshold networks, have been 
developed. The resulting networks have size and complexity comparable to developmental gene regulation modules as
observed in animals, e.g., \emph{Drosophila} \citep{Dassow2000,Albert2003} or \emph{Hydra} \citep{Bosch2003}. 
The threshold networks (as models for transcriptional regulation networks) process position information in
a hierarchical manner; in the present study, hierarchy levels were limited to three, but realizations with more
levels of hierarchy, i.e.\ more ``pre-processing'' of information are also possible. Similar hierarchical and modular organization are
typical signatures of gene regulatory networks in organisms \citep{Davidson2001}. 

Robustness of the model was studied in detail for two types of perturbations, stochastic update errors (noise)
and directed cell flow.  A first order phase transition is observed for vanishing noise and a second order phase transition 
at increasing cell flow. Fluctuations of the noise-controlled boundary position were studied numerically for finite size
systems and analytically in the continuum limit. We find that the relative size of fluctuations vanishes with $1/N_C$,
which means that the boundary position becomes sharp in the limit of large system sizes. 
This means that, based on the proposed local mechanism of
coupled regulatory networks, positional information can be reliably controlled also in large tissues,
which is problematic in the alternative case of morphogenetic gradients that
are typically limited to relatively small spatial domains.
Dynamics under cell flow
was studied in detail numerically and analytically by a mean field approximation. A basic observation is that 
noise-induced perturbations act as quasi-particles that 
stabilize the pattern against the directed force of cell flow.
Hence, we make the interesting observation that noise in local gene expression states 
(over several orders of magnitude in the relevant dynamical
parameters) contributes to robustness of the global developmental dynamics;
furthermore, this is a truly emergent property of the spatial system, which was not selected
during simulated evolution.
At a critical cell flow rate, there is a second order
phase transition towards a vanishing domain size or a 
domain extending over the whole system, depending on the direction
of cell flow, respectively. 
The proposed local mechanism of developmental pattern control also works in growing tissues,
reproduces pattern regeneration after cutting a tissue in half, and is robust
against noise in the recognition of the body axis direction.

Let us briefly compare the prospects and limitations our model with respect to other recent models suggested for pattern formation, 
and with experimental evidence. "Local" models of pattern formation, in contrast to older models
that require long-range diffusion (which is problematic in multi-cellular environments in a number of regards),
have been suggested in the context of juxtacrine signalling (JS, \cite{Monk1998,Owen2000}) and homeoprotein intercellular
transfer (HIT, \cite{Kasatkin2007,Holcman2007}). Similar to JS models with relay, traveling waves/excitations emerge
in our model as a means to provide long range communication. Sharpness and precision of boundary regulation is shared
with HIT models, where, however, this property arises from a different mechanism (meeting of morphogenetic gradients).
While in HIT models noise can substantially affect boundary regulation, an essential property of the model
analyzed in our study is its astonishing robustness against noise. When cell movement is present, noise in fact considerably
contributes to pattern regulation and -stabilization.
An evident limitation of the model arises from the fact that it accounts for regulation of sharp expression boundaries,
but not for graded expression patterns.
Sharp boundaries are indeed found for many genes in development (examples in Hydra are Hedgehog \citep{Kaloulis2000} and the sharp basal
border of HyBMP5-8b \citep{Reinhardt2004}),
while other genes such as CnNK2 \citep{Grens1996} and Dkk \citep{Augustin2006} exhibit more graded expression patterns along the body axis.
It seems quite natural to assume that, in addition to local mechanims as proposed in our model, other mechanims of pattern formation
are present in developing organisms that work on other scales and in different functional contexts, involving regulatory
processes based on graded expression profiles. The hierarchical interplay of such diverse regulatory mechanisms might substantially
contribute to the astonishing robustness of developmental processes. Going beyond basal metazoa such as {\em Hydra},
other interesting applications of our model are conceivable. For instance, local communication systems between adjacent cells as, for
example, the  Delta/Notch systems, play a decisive role in vertebrate development, with traveling waves providing long-range
synchronization of developmental processes \citep{Ozbudak2008}.

Several extensions of the model as described in this paper are conceivable. 
In the present model, the cell flow rate $r_f$ is considered as a free parameter, the global pattern, however, can be controlled
easily by an appropriate choice of the error rate $r_e$. This may suggest to extend the model by introduction of some kind of
dynamical coupling between $r_e$ and $r_f$, treating $r_f$ as a function of $r_e$. Interestingly, similar approaches have been
studied by Hogeweg \citep{Hogeweg2000} and Furusawa and Kaneko \citep{Furusawa2000,Furusawa2003}: 
In both models of morphogenesis, the rate of cell divisions is
controlled by cell differentiation and cell-to-cell signaling. Dynamics in both models, however, is deterministic.
An extension of our model as outlined above may open up for interesting studies how {\em stochastic} signaling events could
control and stabilize a global expression pattern and cell flow as an integrated system. Other possible extensions of the model concern
the dimensionality: In two or three dimensions other mechanisms of symmetry breaking might be present, possibly
leading to new, interesting dynamical effects. 

A Java applet simulation of the model can be found at\\
 {\ttfamily http://www.theo-physik.uni-kiel.de/$\sim$rohlf/development.html}.

\section{Acknowledgements}
We thank T.C.G.\ Bosch, T.W.\ Holstein, and U.\ Technau  
for pointing us to current questions in Hydra development. 
T.R.\ acknowledges financial support from 
the Studienstiftung des deutschen Volkes (German National Academic Foundation).


\begin{appendix}

\section{Genetic algorithm searches}
Let us briefly recapitulate here how the rule table of the model has been obtained by the aid of a 
genetic algorithm. 
\subsection{Definition of the GA}
In order to find a set $\cal F$ of update rules that solve the problem as formulated in section II, 
cellular automata have been evolved using genetic algorithms \citep{Mitchell1994}. Genetic algorithms are 
population-based search algorithms, which are inspired by the interplay of random mutations 
and selection as observed in biological evolution \citep{Holland1975}. Starting from a randomly 
generated population of $P$ rule tables $f_n$, the algorithm optimizes possible solutions by
evaluating the fitness function
\begin{eqnarray}
\Phi(f_n) = \frac{1}{(T_u-T)\cdot N_C}\sum_{t = T_u-T}^{T_u}\left(\sum_{i=0}^{[\alpha\cdot N_C]-1}\delta_{\sigma_i^n(t),2} \right. \nonumber\\
        \left. + \sum_{i=[\alpha\cdot N_C]}^{N_C-1}\{1-\delta_{\sigma_i^n(t),2}\} \right).
\end{eqnarray}
The optimization algorithm then is defined as follows: 
\begin{enumerate}
\item Generate a random initial population $\cal F$ $= \{ f_1,...,f_P \}$ of rule tables.
\item Randomly assign system sizes $N_C^{min} \le N_c^n \le N_C^{max}$ to all rule tables.
\item For each rule table, generate a random initial state vector.
\item Randomly mutate one entry of each rule table (generating a population $\cal F^*$ of mutants).
\item Iterate dynamics over $T_u$ time steps for $\cal F$ and  $\cal F^*$. 
\item Evaluate $\Phi(f_n)$ and $\Phi(f_n^*)$ for all rule tables $0 < n \le P$, averaging over the past 
$T_u-T$ update steps (with an additional penalty term
      if $f_n$ does not converge to a fixed point).
\item For each $n$, replace $f_n$ with $f_n^*$, if $\Phi(f_n) \le \Phi(f_n^*)$.
\item Replace the least fit solution by a duplicate of the fittest one.
\item Go back to step 2 and iterate.
\end{enumerate}
The outcome of this search algorithm is a set of rule tables, which then can be ``translated''
into (spatially coupled) Boolean networks or threshold networks with suitable thresholds.
This yields a set of (minimal) dynamical networks which solve the pattern formation
task by means of internal information processing.
\begin{figure}[htb]
\begin{center}
\resizebox{85mm}{!}{\includegraphics{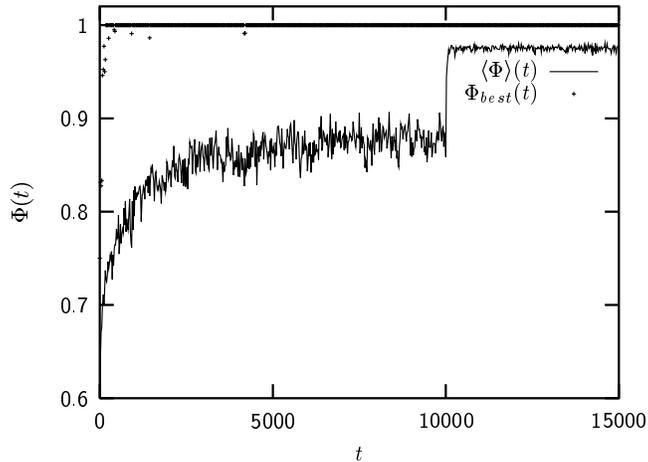}}
\end{center}
\caption{\small Average fitness $\langle \Phi\rangle(t)$ of the mutant population $\cal F^*$ 
and fitness of the highest fitness mutant $\Phi_{best}(t)$ as a function of simulation time 
during the genetic algorithm run that lead to the high fitness solution used in this paper.
At time step 10000 mutations
were turned off, in order to test the established population of optimized rule tables under different
initial conditions (this corresponds to the sharp increase of  $\langle \Phi\rangle(t)$ at time step 10000). 
The evolved population of rule tables has an average fitness of about 0.98, independent from the initial conditions and
system size $N_C$ (in the tested range, i.e. $15 \le N_C \le 150$).}
\label{evo}
\end{figure}
\begin{figure}[htb]
\begin{center}
\resizebox{85mm}{!}{\includegraphics{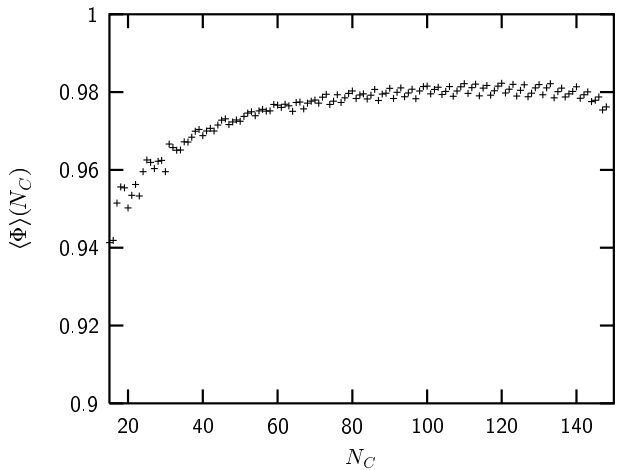}}
\end{center}
\caption{\small Average fitness of the highest fitness rule table as a function of the system size $N_C$. For system sizes
$N_C \ge 80$ the fitness is almost constant at about 0.98. Notice that the decrease of the fitness for small
$N_C$ is an effect of the \emph{dynamics}, not of the genetic algorithm implementation (all $N_C$ in the range $15 \le N_C\le 150$
were tested with equal probability), hence the dynamics of pattern formation may impose a lower boundary
on the range of system (animal) sizes tolerated by natural selection.  }
\label{fitOfN}
\end{figure}
\begin{figure}[htb]
\begin{center}
\resizebox{85mm}{!}{\includegraphics{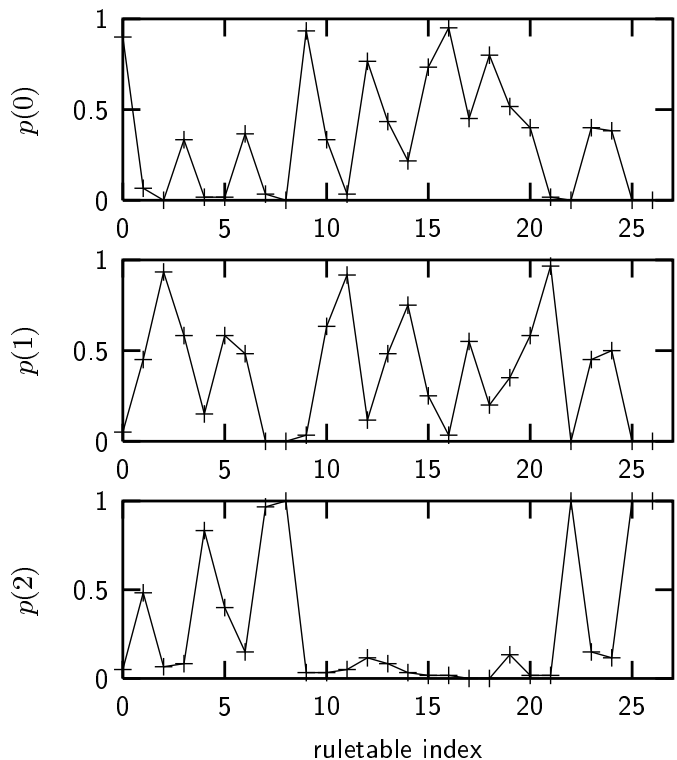}}
\end{center}
\caption{\small Frequency distribution $p(\sigma_i)$ of outputs as a function of the rule table
index as denoted in table 1. Ensemble statistics is taken
over 80 different solutions with $\Phi \ge 0.96$. The upper panel shows the distribution for
$\sigma_i = 0$, the middle panel  the distribution for $\sigma_i = 1$ and the lower panel
 the distribution for $\sigma_i = 2$.}
\label{rulstat}
\end{figure}
\begin{figure}[tb]
\begin{center}
\resizebox{85mm}{!}{\includegraphics{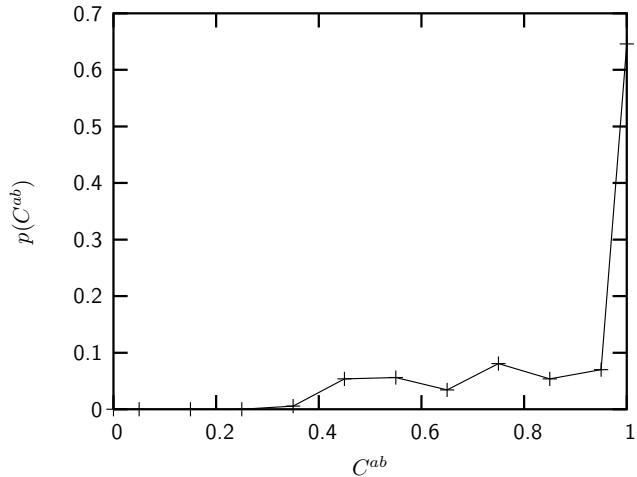}}
\end{center}
\caption{\small Frequency distribution $p(C^{ab})$ of two point correlations $C^{ab}(\sigma,\sigma^{'})$ 
of rule table entries, as defined in Eqn.\ (B2), averaged over all possible pairs of rule table entries. 
About 65\% of rule table entries have correlation 1.0, for the rest the correlation is lower.}
\label{rulcorr}
\end{figure}
\begin{figure}[htb]
\begin{center}
\resizebox{85mm}{!}{\includegraphics{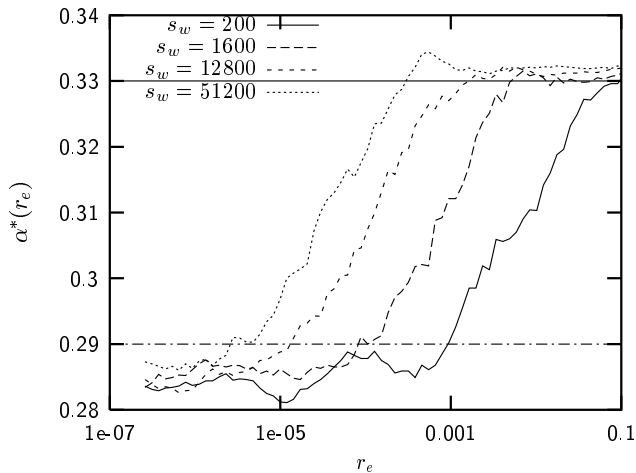}}
\end{center}
\caption{\small Average domain boundary position $\alpha^*$ as a function of the error rate $r_e$, sampled
over update windows of different lengths $s_w$ (ensemble statistics, 400 different initial conditions
for each data point). The abscissa is logarithmic.
With increasing $s_w$, the transition from the solution
$\alpha^*_{det} = 0.281$ under deterministic dynamics to $\alpha^* = 1/3$ under noise 
is shifted towards $r_e = 0$. The two straight lines define a lower boundary $\alpha^*_{low}$ and 
a upper boundary $\alpha^*_{up}$, as explained in the text. }
\label{pt0Shift}
\end{figure}

\begin{figure}[htb]
\begin{center}
\resizebox{85mm}{!}{\includegraphics{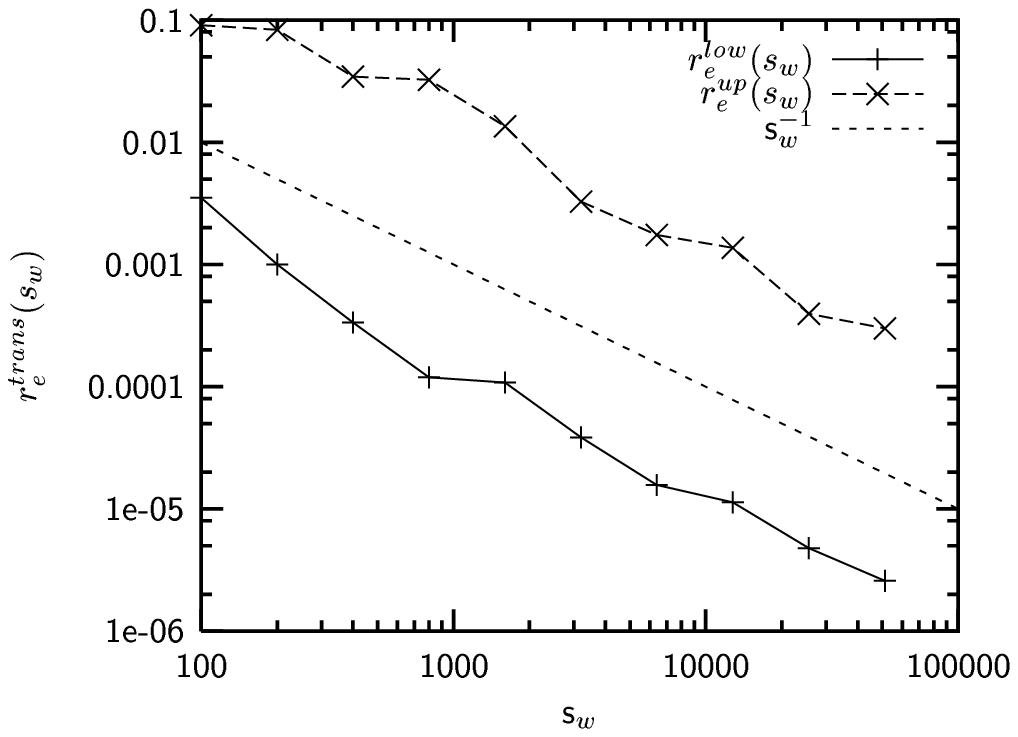}}
\end{center}
\caption{\small Finite size scaling of the upper and lower transition points $r_e^{up}$ and $r_e^{low}$,
i.e. the points where $\alpha^*$ crosses $\alpha^*_{low}$ and $\alpha^*_{up}$, respectively (Fig. \ref{pt0Shift}),
as a function of the sampling window length $s_w$.
Both  $r_e^{up}$ and $r_e^{low}$ vanish $\propto s_w^{-1}$, as indicated by the line
with slope $-1$ in this log-log-plot. }
\label{pt0Scal}
\end{figure}
\subsection{Evolution of cellular automata}

The genetic algorithm sketched above is run with the following parameter
choices: $15 \le N_C \le 150$, i.e.\ during GA runs the system size is 
varied randomly between $15$ and $150$ cells, and the population size is set to
$P = 100$.
 Fig.\ \ref{evo} shows the fitness of the highest fitness mutant and the average
fitness of the population as a function of the number of successive mutation
steps during optimization. A useful solution is found rather quickly (after about
200 updates), with further optimization observed during further 10000 generations. 
At time step 10000 mutations are turned off, thus now
the average fitness of the established population under random initial conditions
and random fluctuations of the system size $N_C$ is tested. The average
fitness $\Phi \approx 0.98$ indicates a surprisingly high robustness
against fluctuations in the initial start pattern, indicating that the system
is capable of \emph{de novo pattern formation}.
In the ``fitness picture'', Fig. \ref{fitOfN} confirms
that the dynamically regulated domain size ratio $\alpha/(1-\alpha)$
indeed is independent of system size (proportion regulation),
there is only a weak decay of the fitness at small values of $N_C$.
Interestingly, one also observes that for regeneration of Hydra polyps
from random cell aggregates a minimum number of cells is required
\citep{Technau2000}. The model suggests that this observation might be explained
by the dynamics of an underlying pattern generating mechanism, i.e.\ that there
has to be a minimum diversity in the initial condition for successful
de novo pattern formation.

\section{Statistical analysis of solutions}

An interesting question is how ``difficult'' it would be for an evolutionary 
process driven by random mutations and selection to find solutions
for the pattern formation problem based on neighbor interactions
between cells. As we showed above, the genetic algorithm finds the correct solution
fast, however, this does not necessarily
mean that biological evolution could access the same solution as fast. If there
is only one, singular solution, evolution may never succeed finding it,
as the genotype which already exists cannot be modified in an arbitrary
way without possibly destroying function of the organism (\emph{developmental
constrains}). To illustrate this point, we generated an ensemble of $N_E = 80$ different solutions
with $\Phi \ge 0.96$ and performed a statistical analysis of the rule table
structure.

As one can see in Fig.\ \ref{rulstat}, some positions in the rule table
are quite fixed, i.e., there is not much variety in the outputs,
whereas other positions are more variable.
We note that a number of rule table positions are {\em a priori} fixed due to the constraints
imposed on dynamics. For example, to support a
stable boundary $...2221000...$ as for the model described, five rules become fixed:
$222 \rightarrow 2$, $221 \rightarrow 2$, $210 \rightarrow 1$, $100 \rightarrow 0$ and $000 \rightarrow 0$
(these rules can be clearly distinguished as pronounced peaks in Fig. \ref{rulstat}).
Consistency with boundary conditions fixes more rules. For the model as described
in this manuscript, i.e. boundary conditions $\sigma_{-1} = \sigma_{N_C} = 0$,
the rule $022\rightarrow 2$ becomes fixed, too.

However, the output frequency
distribution alone does not allow to really judge the ``evolvability''
of the solutions: if there are strong correlations between most of the
rule table entries, evolutionary transitions from one solution to another
would be almost impossible. To check this point,
we studied statistical two point correlations between the rule table entries. The 
probability for finding state $\sigma$
at position $a$ and state $\sigma'$ at rule table position $b$ is given by 
\begin{equation} p^{ab}(\sigma,\sigma') = \frac{1}{N_E}\sum_{n=1}^{N_E}\delta_{\sigma^n(a),\sigma}\cdot\delta_{\sigma^n(b),\sigma'}
,\end{equation}
where $\delta$ is the Kronecker symbol and $n$ runs over the statistical ensemble 
of size $N_E$. The two point correlation between $a$ and $b$ then is defined as 
\begin{equation} C^{ab} = c_1\left( \max_{(\sigma, \sigma')} p^{ab}(\sigma,\sigma') - c_2\right) \end{equation}
with $c_1 = 9/8$ and $c_2 = 1/9$ to obtain a proper normalization with respect to the two limiting cases of equal
probabilities ($p^{ab}(\sigma,\sigma{'}) = 1/9 \quad \forall (\sigma,\sigma{'})$) and $p^{ab}(\sigma,\sigma{'}) = 1$ for 
$\sigma = \tilde\sigma, \quad \sigma^{'} = \tilde\sigma^{'}$
and $p^{ab}(\sigma,\sigma') = 0$ for all other $(\sigma,\sigma')$).
Fig.\ \ref{rulcorr} shows the frequency distribution of $C^{ab}(\sigma,\sigma')$, 
averaged over all possible pairs $(a,b)$. About 65\% of rule table positions 
are strongly correlated ($C^{ab} = 1.0$), the rest shows correlation values
between $0.3$ and $1.0$. Hence, we find that the space of solutions is 
restricted, nevertheless there is variability in several rule table positions. 
To summarize this aspect, the pattern formation mechanism studied in this 
paper shows considerable robustness against rule mutations, however,  
a ``core module'' of rules is always fixed. Interestingly, a similar phenomenon 
is observed in developmental biology: Regulatory modules involved 
in developmental processes often are evolutionarily very conservative,
 i.e., they are shared by almost all animal phyla \citep{Davidson2001}, while 
 morphological variety is created by (few) taxon specific genes 
 \citep{Bosch2003} and rewiring of existing developmental modules.

\section{Numerical analysis of the dynamical transition at $r_e=0$}\label{noitrans_app}
The rate equation of the pattern formation system in the presence of
noise with error rate $r_e$ is given by 
\begin{equation}
 2\alpha^* r_e = (1 - \alpha^*) r_e,
\label{rate}
\end{equation}
i.e. $\alpha^* = 1/3$. Comparison to the equilibrium position in the noiseless case indicates
that the system undergoes a step-like discontinuity with respect to  $\alpha^*$ at $r_e = 0$.
A numerical analysis that considers small variations of $r_e$ close to zero and averages
over time windows of variable length $s_w$ can be applied for supporting numerical evidence
and {\em finite size scaling}.

Figs.\ \ref{pt0Shift} and \ref{pt0Scal} show noise 
dependence and finite size scaling of the transition
from the unperturbed solution to the solution under noise.
Considering update time windows of different length $s_w$,
in case of a discontinuity (i.e., a step-like 'jump' of 
the order parameter) at $r_e = 0$ we would expect a shift 
of the transition point $r_e^{trans}(s_w)$ towards $r_e = 0$ 
which is proportional to $s_w^{-1}$ as well as a divergence 
of the slope at the transition point when $s_w$ is increased,  
i.e. $d\alpha^*/dr_e(r_e^{trans}) \to \infty$ when $s_w \to \infty$.

The shift of $r_e^{trans}(s_w)$ is most easily measured by defining a lower and a upper boundary
$\alpha^*_{low}$ and $\alpha^*_{up}$, respectively (Fig. \ref{pt0Shift}); when  $\alpha^*$ crosses these boundaries,
two transition points $r_e^{up}$ and $r_e^{low}$ are obtained.
We find that $r_e^{up} \approx c_{up}s_w^{-1}$ and  $r_e^{low} \approx c_{low}s_w^{-1}$ with $c_{up} > c_{low}$ as
expected (Fig. \ref{pt0Scal}), which implies that the difference $\Delta r_e^{trans}(s_w) := r_e^{up} - r_e^{low}$ scales as
\begin{equation} \Delta r_e^{trans}(s_w) = (c_{up} - c_{low})\,s_w^{-1},\label{delta1} \end{equation} 
hence, because $\Delta \alpha^*(r_e^{trans}) = const. = \alpha^*_{up}- \alpha^*_{low}$,
indeed  $d\alpha^*/dr_e(r_e^{trans})$ diverges when the sampling window size goes to infinity. 

\section{Derivation of stability conditions for growing systems}
\subsection{Symmetric growth at the system boundaries}\label{symmgroapp}
Let us assume we start with a system of $N_0$ cells, with an initial boundary position at cell $N_1$. In the deterministic case $r_e=0$, it is straight-forward to
see that the time dependence of the average boundary position is given by

 \begin{equation} 
 \alpha^*(t) = \frac{N_1 + \frac{1}{2}r_g\,t}{N_0+r_g\,t},
 \end{equation}
hence we have
\begin{equation} 
 \alpha^{\infty} = \lim_{t\to\infty} \alpha^*(t) = \frac{1}{2}.
 \end{equation}

In the case $r_e > 0$, the time dependence of the average boundary position is given by
\begin{equation} 
 \alpha^*(t) = \frac{N_1 + \frac{1}{2}r_g\,t - 2\alpha^*(t)r_e t + (1-\alpha^*(t))r_e t }{N_0+r_g\,t},
 \end{equation}
 which simplifies to
 \begin{equation} 
 \alpha^*(t) = \frac{N_1 + \frac{1}{2}r_g\,t + r_e t }{N_0+r_g\,t + 3r_e t}.
 \end{equation}
 Assuming infinite growth, this leads to the asymptotic boundary position
 \begin{equation} 
 \alpha^{\infty} = \lim_{t\to\infty} \alpha^*(t) = \frac{\frac{1}{2}r_g + r_e   }{r_g+3r_e}.
 \end{equation}
 Fig. \ref{growth} shows $\alpha^{\infty}(r_g)$ for four different values of $r_e$; it becomes evident that an approximately 'correct'
 proportion regulation requires $r_e$ to be at the order of $r_g$ or larger, i.e. $r_e/r_g \ge 1$. While $r_e$
 (the rate of regulatory signals) may not be increased significantly above the growth rate $r_g$, due to metabolic constraints,
 in later stages of development the steady decrease of $r_g$ will ensure that the condition $r_e/r_g \ge 1$ is fullfilled and proportion regulation
 approaches the steady state of the adult organism.

\subsection{Growth by homogeneous cell proliferation}\label{homgroapp}
We require that de-novo pattern formation has taken place in a system of stationary size $N_0$ and has converged
to its final pattern.

Assuming homogeneous cell proliferation with a probability $p_d$ per cell, it is easy to see that proliferation of "blue"
($\sigma_i = 2$) and "black" ($\sigma_i = 0$) cells conserves pattern proportions, proliferation of the "red" boundary 
cell ($\sigma_i = 1$), however, leads to readjustment of the boundary at its original position before proliferation,
and hence slightly reduces $\alpha$ (see inset of Fig.\ref{homgrowth_fig}). 
Since, in this case, boundary readjustment needs two update time steps and occurs
with probability $p_d$, $\alpha(t)$ is given by
\begin{equation}
\alpha(t) = \frac{N_1(t-1)(1+p_d)-p_d/2}{N(t)}.\label{alphahomgrowth_eq}
\end{equation}
System size grows geometrically, i.e. $N(t) = N_0(1+p_d)^t$. Inserting this dependence into Eq. (\ref{alphahomgrowth_eq})
and using $N_1(t-1) = \alpha(t-1)N(t-1)$, 
it follows that
\begin{equation}
\alpha(t) = \alpha(t-1) - \frac{p_d}{2N_0(1+p_d)^t}.
\end{equation}
Recursively inserting for $\alpha(t-\tau)$ for $\tau \in \{1,...,t-1\}$, we conclude that
$\alpha(t)$ is given by
\begin{equation}
\alpha(t) = \alpha_0 - \frac{p_d}{2N_0}\sum_{\tau=1}^{t}(1+p_d)^{-\tau} = \alpha_0 -\frac{1}{2N_0}\left\{ 1-(1+p_d)^{-t}  \right\}
\end{equation}
This implies that the asymptotic boundary position in the limit $t \to \infty$ is independent from $p_d$:
\begin{equation}
\alpha_{\infty} = \alpha_0 - \frac{1}{2N_0}
\end{equation}
Hence deviations at the order of $1/N_0$ from the boundary position 
$\alpha_0 = 0.281$ of non-growing systems are found.

For the case when noise is present, it is not possible to find a general solution since proliferation
can affect both the velocity and the type of particles travelling through the domains in intricate ways
(in the case of symmetric growth at the system boundaries,
as discussed in the previous subsection, this problem is avoided). If $p_d$ is very small,
however, we can assume that proliferation events and particle propagation are essentially decoupled
and that the system has enough time to relax to a stationary state between proliferation events
\footnote{Furthermore, one can show that single proliferation events that occur in or near quasi-particles
have the same effect as certain subclasses of one-site errors as discussed in section \ref{noise_sec}. From this
we conclude that - for moderate $p_d$ -  the statistics of boundary readjustments remains unchanged,
just as it was found in section \ref{noise_sec} for error rates $r_e < 1/2$.}. In this limit,
we can generalize Eqn. (\ref{rate}) in a straight-forward way:
\begin{equation}
2\alpha r_e + p_d = (1-\alpha)r_e,
\end{equation} 
leading to
\begin{equation}
\alpha = \frac{1}{3}\left(1 - \frac{p_d}{r_e}\right).
\end{equation}
$r_e = pN$ is a monotonously growing function in time for fixed error probability $p$, hence
it follows that, for large $t$, the boundary position converges to the same value $\alpha = 1/3$ as for constant size
systems.

\end{appendix}

\bibliographystyle{elsarticle-harv}
\bibliography{rohlf_bornholdt_refs_jtb}

\end{document}